\documentclass{jpp}
\usepackage{graphicx}
\usepackage{hyperref}
\usepackage{color}

\usepackage[utf8]{inputenc}
\usepackage[T1]{fontenc}
\usepackage{amsmath}

\newcommand{\vect}[1]{\mbox{\boldmath $#1$}}

\newcommand{\changed}[1]{{#1}}
\newcommand{\DMerc}{D_{\mathrm{Merc}}}
\newcommand{\nfp}{n_{\mathrm{fp}}}
\newcommand{\etabar}{\bar{\eta}}

\shorttitle{Mapping the space of quasisymmetric stellarators
}
\shortauthor{Matt Landreman
}

\title{Mapping the space of quasisymmetric stellarators using optimized near-axis expansion}

\author{Matt Landreman\aff{1}
  \corresp{\email{mattland@umd.edu}},
 }

\affiliation{\aff{1}Institute for Research in Electronics and Applied Physics, University of Maryland,
College Park MD, 20742, USA
}

\begin{document}

\maketitle

\begin{abstract}
A method is demonstrated to 
rapidly calculate the shapes and properties of
quasi-axisymmetric and quasi-helically symmetric stellarators.
In this approach, optimization is applied to the equations of magnetohydrodynamic equilibrium and quasisymmetry, expanded in the small distance from the magnetic axis, as formulated by Garren and Boozer [\emph{Phys. Fluids B}, 3, 2805 (1991)].
Due to the reduction of the equations by the expansion, the computational cost is significantly reduced, to times on the order of 1 cpu-second, enabling wide and high-resolution scans over parameter space.
In contrast to traditional stellarator optimization, here the cost function serves to maximize the volume in which the expansion is accurate.
A key term in the cost function is $|| \nabla\mathbf B ||$,
to maximize scale lengths in the field.
Using this method, a database of $5\times 10^5$ optimized configurations is calculated and presented.
Quasisymmetric configurations are observed to exist in continuous bands, varying in the ratio of the magnetic axis length to average major radius.
Several qualitatively new types of configuration are found, including 
quasi-helically symmetric fields in which the number of field periods is two or more than six.
\end{abstract}

\section{Introduction}

Stellarators can potentially provide steady-state plasma confinement with minimal recirculating power, passive stability, and no danger of disruptions. However stellarators require careful shaping of the field in order to confine trapped particles. This optimization is challenging because the space of plasma shapes is high-dimensional and known to contain multiple local minima \citep{Bader2019}. Numerical optimization with local optimization algorithms is effective at finding individual configurations, but it does not provide a global picture of the space of solutions. Global optimization is difficult due to the high number of dimensions, and the dimensionality also makes parameter scans over the full space of possible shapes infeasible. Due to these challenges, it is not clear that all the interesting regions of parameter space have been found.

In this work, we attempt a global view of the space of optimized stellarators by using approximate magnetohydrodynamic (MHD) equilibria instead of full 3D equilibria, greatly reducing computational cost. In particular we will use an expansion about the magnetic axis \citep{Mercier1964,Solovev,GB1}, \changed{a closed field line representing the innermost flux surface}. This expansion reduces the 3D partial differential equations of MHD equilibrium to 1D ordinary differential equations in the toroidal direction, lowering the time required to compute and diagnose a configuration by several orders of magnitude. It then becomes feasible to carry out high-resolution multi-dimensional parameter scans, resulting in large databases of stellarator configurations. 
While the expansion is approximate, it is necessarily accurate in the core (out to some minor radius) of any stellarator, even one for which the aspect ratio of the plasma boundary is low.

In this work we focus on the condition of quasisymmetry, one effective strategy for confining trapped particles \citep{Boozer1983,NuhrenbergZille,HelanderReview}. Quasisymmetry is a condition that the magnitude $B=|\vect{B}|$ of the magnetic field $\vect{B}$ is effectively 2D instead of 3D, with the continuous symmetry providing a conserved quantity that ensures confinement. Two types of quasisymmetry are possible near the axis: quasi-axisymmetry (QA), $B=B(r,\theta)$, and quasi-helical symmetry (QH), $B=B(r, \theta - N\varphi)$. Here $r$ is a flux surface label, $(\theta,\varphi)$ are the Boozer poloidal and toroidal angles, and $N$ is an integer.
Although the weaker condition of omnigenity may be sufficient for trapped particle confinement, we focus here on quasisymmetry because the condition is easier to express mathematically,
and since the omnigenous generalizations of QA and QH provide no extra freedom near the magnetic axis \citep{PaperIII}.

In the first few orders of the near-axis expansion, it is possible to impose quasisymmetry directly, without optimization. Yet optimization is still useful, to maximize the minor radius over which the expansion is accurate. For most parameters of the near-axis model (which include the axis shape and a few other numbers), the minor radius over which the expansion is accurate is quite small. Therefore the next order terms in which quasisymmetry is broken are significant unless the plasma has extremely high aspect ratio, $> 10$. By optimizing the parameters of the near-axis model, configurations can be obtained for which the expansion is accurate even at lower aspect ratios, in the range 5-10, typical of stellarator experiments. These configurations then have quasisymmetry over a significant volume.
Since quasisymmetry is necessarily broken at third order in the expansion \citep{GB1,GB2}, having good quasisymmetry over a large volume probably requires that the plasma be accurately described by the lower orders of the expansion \citep{RodriguezThesis}.

The method for generating stellarator configurations in this work is complementary to traditional stellarator optimization, in which the boundary shape of a finite-aspect-ratio plasma is the parameter space, and a fully 3D MHD equilibrium code is run to evaluate the objective function. The method in this paper is necessarily approximate, but wider surveys over parameters are feasible. Conventional optimization is more accurate, but global optimization is more difficult. The two approaches could be used together, with the near-axis method identifying rough configurations that could be passed as an initial condition to conventional optimization for refinement.

Optimization has been applied to near-axis expansions in several previous publications. In \cite{r2GarrenBoozer}, some results were shown from a preliminary version of the approach here, but the optimization method was not explained in detail. One purpose of the present article is to give a detailed presentation. 
A different approach for choosing parameters of the near-axis model and mapping the space of solutions was proposed in \cite{RodriguezLandscape}.
Optimization of a near-axis quasi-isodynamic (QI) stellarator was presented recently in \cite{JorgeQI}. Other optimizations of QI near-axis parameters are shown in \cite{Mata}.

The near-axis expansion for quasisymmetry has been discussed in detail in previous publications, but a brief review is given in section \ref{sec:expansion}. In section \ref{sec:opt}, we describe the optimization problem for expanding the minor radius over which the expansion is accurate, and for achieving other desired physics properties. Next, a wide scan over parameters is presented in section \ref{sec:scan}, and the space of QA and QH configurations obtained is discussed. A few examples of configurations found in the scan are shown in section \ref{sec:configs}. We discuss the results and conclude in section \ref{sec:conclusions}.


\section{Garren-Boozer expansion and diagnostics}
\label{sec:expansion}

Here we give an overview of the near-axis expansion used for optimization, highlighting the quantities that are inputs and outputs for each stellarator configuration.
We use the form of the expansion introduced by \cite{GB1, GB2}, in which the independent variables are Boozer coordinates.
A detailed discussion can also be found in
\cite{r2GarrenBoozer}. This expansion has also been discussed in
\cite{LandremanMercier,FiguresOfMerit}. There are other ways to carry out expansion about the axis
in which the independent variables are not flux coordinates \citep{Mercier1964,Solovev,Jorge2021}, which will not be considered here.

One input to the near-axis equations is the magnetic axis's shape. The position vector along the axis $\vect{r}_0$ can be expressed as a function of the arclength $\ell$ along the curve. At each point on the axis, the Frenet frame is defined by
\begin{align}
\frac{d\vect{r}_0}{d\ell} = \vect{t}, 
\hspace{0.3in}
\frac{d\vect{t}}{d\ell} = \kappa \vect{n}, 
\hspace{0.3in}
\frac{d\vect{n}}{d\ell} = -\kappa \vect{t} + \tau \vect{b}, 
\hspace{0.3in}
\frac{d\vect{b}}{d\ell} = -\tau \vect{n}.
\label{eq:Frenet}
\end{align}
Here, $(\vect{t},\vect{n},\vect{b})$ are the tangent, normal, and binormal, a set of orthonormal vectors satisfying
$\vect{t}\times\vect{n}=\vect{b}$.
Also, $\kappa$ is the axis curvature, and $\tau$ is the axis torsion. It can be shown that for quasisymmetric configurations, $\kappa$ does not vanish, so the Frenet frame is well behaved.

The position vector $\vect{r}$ of a general point (not necessarily on the axis) can then be written
\begin{align}
\label{eq:positionVector}
\vect{r}(r,\vartheta,\varphi) = \vect{r}_0(\varphi)
+X(r,\vartheta,\varphi) \vect{n}(\varphi)
+Y(r,\vartheta,\varphi) \vect{b}(\varphi)
+Z(r,\vartheta,\varphi) \vect{t}(\varphi),
\end{align}
where $r$ is a minor radius coordinate, $\varphi$ is a toroidal angle, and $\vartheta$ is another coordinate. We specifically define these three coordinates as follows. Letting $\psi$ denote the toroidal flux divided by $2\pi$, an effective minor radius $r$ can be defined via $2\pi|\psi| =  \pi r^2 B_0$, where 
 $B_0>0$ is the magnetic field strength on the axis, which is constant in quasisymmetry. Note that $r$ is a flux function, and not identical to the Euclidean distance to the axis or the magnitude of $\vect{r}$. We employ the poloidal and toroidal Boozer angles $\theta$ and $\varphi$, in terms of which the field is
\begin{align}
\label{eq:BoozerCoords}
\vect{B} = &\nabla\psi \times\nabla\theta + \iota \nabla\varphi \times\nabla\psi, \\
 = &\beta \nabla\psi + I \nabla\theta + G \nabla\varphi, \nonumber
\end{align}
where $I$ and $G$ are constant on flux surfaces. The remaining coordinate in (\ref{eq:positionVector}) is defined as $\vartheta = \theta - N \varphi$, where $N$ is a constant integer, making $\vartheta$ a poloidal or helical angle for $N=0$ and $N \ne 0$ respectively. Defining $\iota_N = \iota - N$, then
\begin{align}
\vect{B} = &\nabla\psi \times\nabla\vartheta + \iota_N \nabla\varphi \times\nabla\psi,
\label{eq:straight_field_lines_h}
\\
 =& \beta \nabla\psi + I \nabla\vartheta + (G+NI) \nabla\varphi.
\label{eq:Boozer_h}
\end{align}

We now consider $r$ to be small compared to length scales associated with the axis. In this case we can expand $X$, $Y$, and $Z$ in (\ref{eq:positionVector}) as
\begin{align}
X(r,\vartheta,\varphi)
= r X_1(\vartheta,\varphi) + r^2 X_2(\vartheta,\varphi) + r^3 X_3(\vartheta,\varphi) + \ldots.
\label{eq:radial_expansion}
\end{align}
Similar expansions hold for $Y$ and $Z$.
The field strength $B$ and coefficient $\beta$ can be expanded in the same way but with an $r^0$ term:
\begin{align}
\label{eq:radial_expansion_B}
B(r,\vartheta,\varphi)
= B_0(\varphi) + r B_1(\vartheta,\varphi) + r^2 B_2(\vartheta,\varphi)+ r^3 B_3(\vartheta,\varphi) + \ldots.
\end{align}
Flux functions ($\iota$, $G$, $I$, and the pressure $p$) must be even with respect to $r$ and so their expansions contain only even powers of $r$:
\begin{align}
p(r) = p_0 + r^2 p_2 + r^4 p_4 + \ldots.
\end{align}
The profile $I(r)$ is proportional to the toroidal current inside the flux surface, so $I_0 = 0$.

Considerations of analyticity near the magnetic axis imply that poloidally varying quantities must have the form
\begin{align}
\label{eq:poloidal_expansions}
B_1(\vartheta,\varphi) = &B_{1s}(\varphi) \sin(\vartheta) + B_{1c}(\varphi) \cos(\vartheta), \\
B_2(\vartheta,\varphi) = &B_{20}(\varphi) + B_{2s}(\varphi) \sin(2\vartheta) + B_{2c}(\varphi) \cos(2\vartheta). \nonumber
\end{align}
(For a more detailed argument see appendix A of \cite{PaperI}.)
This same form applies also to $X$, $Y$, $Z$, and $\beta$.

So far, the position vector has been expressed as a power series in $r$. Taking derivatives of this position vector with respect to the three coordinates, the dual relations \changed{\citep{Dhaeseleer}} can then be used to evaluate $\nabla \psi$, $\nabla\vartheta$, and $\nabla\varphi$. The results are substituted into (\ref{eq:straight_field_lines_h})-(\ref{eq:Boozer_h}). Equating these covariant and contravariant forms of $\vect{B}$, powers of $r$ can be collected at each order. Moreover, the inner product  (\ref{eq:straight_field_lines_h})$\cdot$(\ref{eq:Boozer_h}) yields an expression for the field strength, $B^2/(G+\iota I) = \nabla\psi\cdot\nabla\vartheta\times\nabla\varphi$, providing an additional equation at each order in $r$. One more equation is provided by MHD equilibrium, $(\nabla\times\vect{B})\times\vect{B} = \mu_0\nabla p$. Here, only the $\nabla\psi$ component provides new information. Finally, if quasisymmetry is desired, the condition $B = B(\psi, \vartheta)$ can be imposed. These conditions provide an increasing number of constraints at each order in $r$.

The conditions obtained at each order are now summarized, for the case of quasisymmetry.
At leading order, $B_0$ is independent of $\varphi$. It is also determined that $\varphi = 2\pi \ell / L$ and $|G_0|=B_0 L/(2\pi)$, where $L$ is the axis length. While $B_0$ can be considered an input parameter, it merely scales the field strength of the configuration and so does not provide true flexibility. At next order, $B_{1s}$ can be set to zero using the freedom in the origin of $\vartheta$, and $B_{1c}$ must be independent of $\varphi$. Following \cite{GB1} we use the constant $\etabar=B_{1c}/B_0$. The quantity $\etabar$ thus controls how much $B$ varies on a flux surface of given minor radius, via
\begin{equation}
B = B_0 \left[ 1 + r \etabar \cos \vartheta + O(r^2) \right].
\label{eq:modB}
\end{equation}
We consider $\etabar$ to be another input of the calculation.
\changed{Also, it can be shown that $N$ must equal the number of times the axis normal vector rotates poloidally about the axis as the axis is traversed toroidally. Typically $|N|$ equals either 0 or the  number of field periods $\nfp$, as is the case for all configurations found in this work, though other values of $N$ are allowed as well.}

A key constraint at this order is a Ricatti equation, eq (2.14) in \cite{r2GarrenBoozer}, an ordinary differential equation (ODE) in $\varphi$. This equation relates $\kappa$, $\tau$, $\iota_0$, $I_2$, \changed{$\etabar$}, and the $O(r)$ flux surface shapes. As discussed in the appendix of \cite{PaperII}, it is convenient to consider as inputs $I_2$ and the deviation from stellarator symmetry at $\varphi=0$, in which case there is a unique solution for $\iota_0$ and the first-order surface shape. 
\changed{
Alternatively, $\iota_0$ could be considered the input and $I_2$ the output \citep{Rodriguez2022weakly}, but we will not do this here.}
For all work in this paper we assume stellarator symmetry and no current density on the axis, $I_2=0$.
\changed{(To include $I_2$ and non-stellarator-symmetric configurations, no substantial changes to the methods in this paper would be required.)}
Therefore there are no inputs to the model other than $\etabar$ at this order. 
At this order, the flux surface shapes are rotated ellipses \changed{(in the plane perpendicular to the magnetic axis)} centered on the axis. \changed{Generally the elongation varies with $\varphi$.}

Proceeding to next order in $r$, it was found by \cite{GB1} that it is not possible to fully specify $B_2$ for a general axis shape, meaning it is not possible to achieve quasisymmetry at this order for most axis shapes. To handle this complication we proceed as in \cite{r2GarrenBoozer}, only partially imposing quasisymmetry at this order. Specifically, we treat $B_{2c}$ and $B_{2s}$ as inputs, but consider $B_{20}$ an output. For stellarator symmetry, $B_{2s}=0$, so this quantity will not be considered further here. For quasisymmetry, $B_{2c}$ is constant (independent of $\varphi$), providing one more scalar input parameter. Then, $B_{20}(\varphi)$ \changed{can be} computed from a linear system of ODEs. These equations and the surface shapes depend on $p_2$, representing the leading behavior of the pressure near the axis. The flux surface shapes at this order include triangularity and Shafranov shift.

It is possible to consider higher order terms in the expansion. However we stop here, at $O(r^2)$, for all results in this paper. This order is sufficient for representing realistic stellarator shapes. If one were to proceed to higher order, quasisymmetry cannot be fully imposed, and choices would need to be made for other functions of toroidal angle, which significantly increases the number of parameters in the model.
\changed{Note also that in an asymptotic expansion such as this one, including higher order terms may decrease rather than increase accuracy.}

To summarize, at the order of interest, the inputs to the near-axis equations are the shape of the axis and the three scalar parameters $\etabar$, $B_{2c}$, and $p_2$. The outputs of the model include $\iota_0$, $B_{20}$, and a parameterization of all flux surface shapes in a neighborhood of the axis.
An efficient numerical method for solving the ODEs to this order is detailed in section 4.2 of \cite{r2GarrenBoozer}, which we also adopt in this work.
Since the shapes of the flux surfaces are known, any one surface can be used as the input to a standard fixed-boundary 3D MHD equilibrium calculation that does not make a near-axis expansion.
From the result, other standard stellarator codes can be run to check the accuracy of the near-axis approximations and to evaluate other physics properties.
Examples of this procedure can be seen in \cite{PaperII, r2GarrenBoozer, JorgeQI}.

Once these near-axis equations are solved for the configuration geometry, some quantities of interest are known immediately, such as $\iota_0$ and the quasisymmetry error associated with variation of $B_{20}$. Many other quantities of interest can be computed directly from the solution at negligible computational cost. One example \citep{LandremanMercier} is the vacuum magnetic well $d^2 V /d \psi^2$, where $V(\psi)$ is the flux surface volume. Another is the Mercier stability criterion $\DMerc$. 

\changed{
From a solution of the near-axis equations, it is also possible to directly compute all the geometric quantities appearing in the gyrokinetic equation and the MHD ballooning equation \citep{JorgeGKGeometry}. However this information will not be exploited here.
}

Other quantities that can be computed include measures for the minor radius over which the expansion is accurate. A precise measure of this radius has not yet been decisively identified, but several estimates have been suggested. Here we will use three estimates. The first two of these are scale lengths associated with the first and second derivatives of the magnetic field vector \citep{FiguresOfMerit}:
\begin{align}
\label{eq:L_grad_B}
L_{\nabla B} &= B \sqrt{2 / ||\nabla\vect{B}||^2},
\\
\label{eq:L_grad_grad_B}
L_{\nabla\nabla B} &= \sqrt{4B / ||\nabla\nabla\vect{B}||
},
\end{align}
Here, $|| \ldots ||$ indicates the square root of the sum of the squares of the elements of the matrix or tensor. In the case of a matrix this is the Frobenius norm. The quantities $L_{\nabla B}$ and $L_{\nabla \nabla B}$ have dimensions of length, and are each normalized so that in the case of an infinite straight wire, they give the distance to the wire. The near-axis expansion is expected to be accurate only if the distance to the axis is small compared to scale lengths in the magnetic field, i.e. for $r \ll L_{\nabla B}$ and $r \ll L_{\nabla\nabla B}$. Therefore it is desirable to maximize these two quantities.
Another estimate for the radius over which the expansion is accurate is $r_c$, defined in section 4 of \citep{FiguresOfMerit}. This quantity is the maximum minor radius at which the second-order flux surface shapes are no longer smooth and nested. The near-axis expansion has necessarily broken down when $r$ is as large as $r_c$, so $r_c$ is a natural target for maximization.

This near-axis model has a limitation related to the bootstrap current. To the order in $r$ considered here, the current profile has only a single degree of freedom, $I_2$, corresponding to a current density that is independent of $r$. However, realistic bootstrap current profiles are peaked at mid-radius, going to zero on axis where the pressure gradient vanishes, and also becoming small at the plasma edge where the collisionality becomes large. Therefore it is not possible to represent realistic profile shapes of bootstrap current in the near-axis model used here. Throughout this paper we proceed by choosing $I_2=0$, consistent with the bootstrap current vanishing on the magnetic axis. However important questions for future research are whether the near-axis model can be extended to higher order to incorporate realistic current profile shapes, and whether it is a reasonable approximation to make $I_2$ equal to a radial average of the current profile.


\section{Optimization problem}
\label{sec:opt}

\begin{figure}
\centerline{
\includegraphics[width=2.5in]{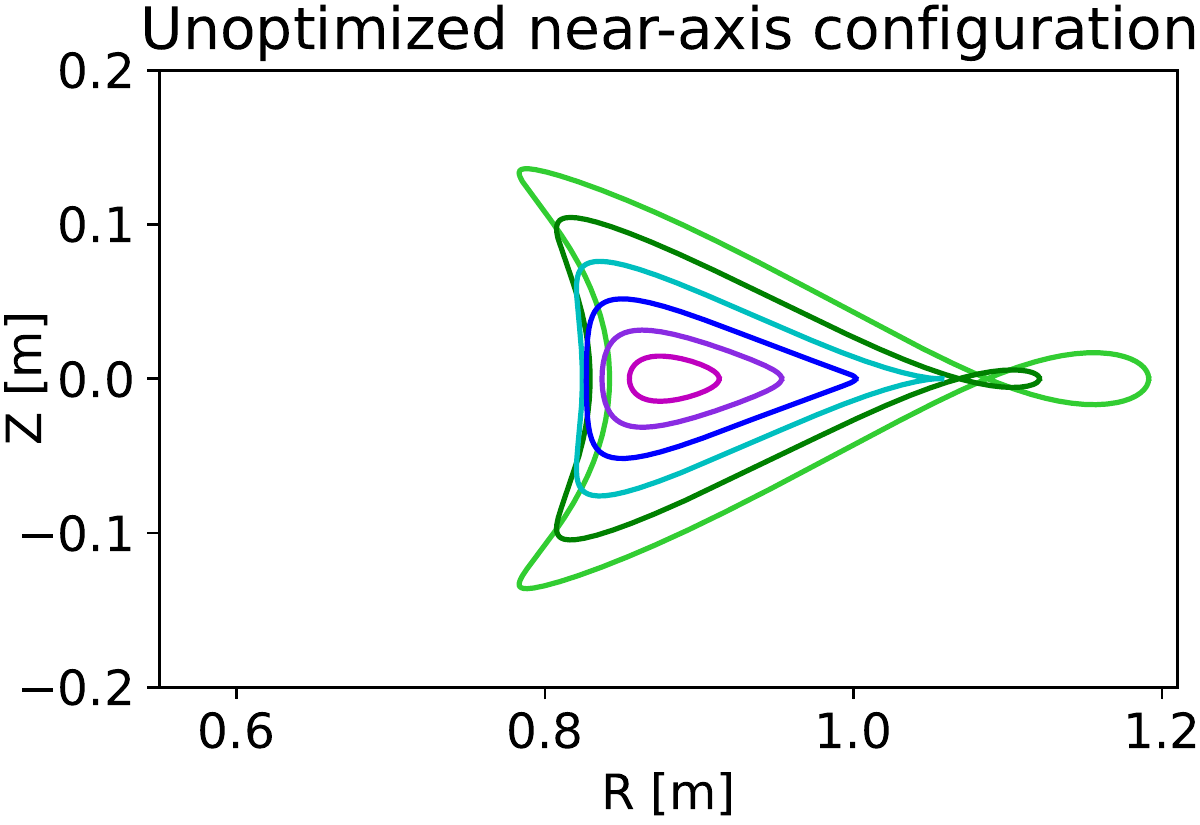}
\hspace{0.15in}
\includegraphics[width=2.5in]{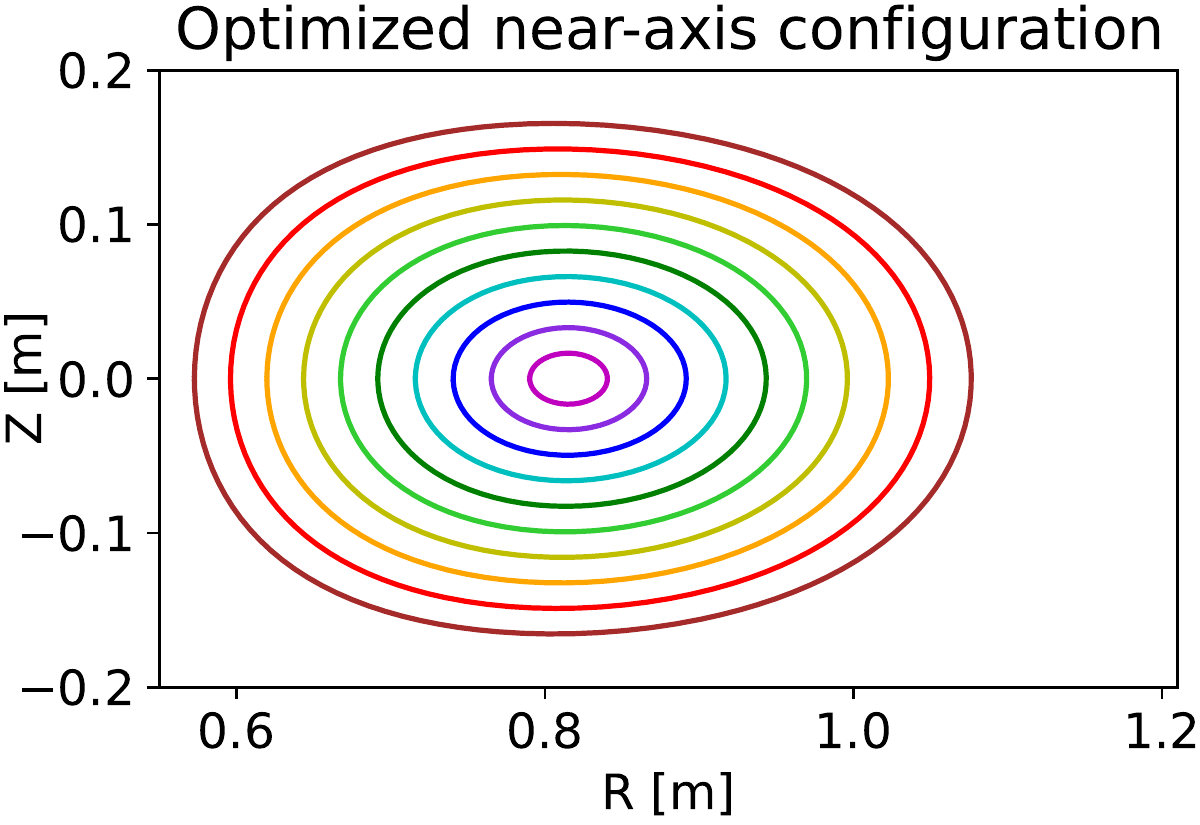}
}
\caption{Left: Generic $O(r^2)$ near-axis configurations tend to be limited to very high aspect ratio before the predicted surface shapes are self-intersecting or non-nested. Right: By optimizing the axis shape, $\etabar$, and $B_{2c}$, the volume of smooth nested surfaces can be dramatically increased. Surfaces shown are  $r=0.02, 0.04, \ldots, 0.12$ m on the left and $r=0.02, 0.04, \ldots, 0.2$ m on the right.
\label{fig:motivation}}
\end{figure}

The motivation for optimization of the near-axis parameters can be seen in figure \ref{fig:motivation}.
The left panel shows a cross-section of flux surfaces computed by the near-axis model for unoptimized input parameters: an axis shape $R(\phi)=1-0.12\cos(2\phi)$ and $Z(\phi)=0.12\cos(2\phi)$, $\etabar=-0.7$, and $B_{2c}=-0.5$. It can be seen that the region of smooth and nested flux surfaces is small, limiting the configuration to very high aspect ratio. For comparison, the right panel of figure  \ref{fig:motivation} shows an optimized near-axis configuration on the same scale. (This configuration will be described in detail in section \ref{sec:QA}). It can be seen that the region of smooth nested surfaces extends to much lower aspect ratio, beyond the range shown. Thus, although quasisymmetry can be imposed directly in the near-axis equations
\changed{(to $O(r)$)}, optimization is still valuable.
\changed{Optimization is also useful for achieving quasisymmetry fully through $O(r^2)$ since, as mentioned above, for general input parameters $B_{20}$ will depend on $\varphi$.}

Let us now present the details of an optimization problem that is effective for the near-axis quasisymmetry equations. The axis shape is represented in cylindrical coordinates $(R,\phi,Z)$ using finite Fourier series,
\begin{eqnarray}
R(\phi) = \sum_{n=0}^{N_F} R_n \cos(\nfp n\phi), \hspace{0.5in}
Z(\phi) = \sum_{n=1}^{N_F} Z_n \sin(\nfp n\phi),
\label{eq:axis_shape}
\end{eqnarray}
where $\nfp$ is the number of field periods, and a finite maximum Fourier number $N_F$ has been chosen. Stellarator symmetry has been assumed. 
\changed{Some axis shapes cannot be represented using (\ref{eq:axis_shape}):
those for which $\phi$ is not monotonic, or those which encircle the $Z$ axis more than once. However
the choice (\ref{eq:axis_shape}) describes every
stellarator experiment to date and so}
is convenient for this initial study.
The parameter space for optimization consists of $\{R_n, Z_n, \etabar, B_{2c} \}$.
The mode $R_0$ is set to 1 and excluded from the parameter space so that the average major radius is held fixed.

The objective function considered is a sum of terms:
\begin{equation}
\label{eq:objective}
f = 
w_{L} f_{L}
+ w_{\iota} f_{\iota}
+ w_{\nabla } f_{\nabla } 
+ w_{\nabla \nabla } f_{\nabla \nabla }
+ w_{B2} f_{B2}
+ w_{\mathrm{well}} f_{\mathrm{well}}
+ w_{\mathrm{Merc}} f_{\mathrm{Merc}},
\end{equation}
where the scalars $w_j$ are weights used to vary the emphasis on the different terms. The individual terms are
\begin{eqnarray}
f_L &=& (L - L_*)^2, \\
f_{\iota} &=& (\iota_0 - \iota_*)^2, \\
\label{eq:f_grad}
f_{\nabla} &=& \frac{1}{L} \int d\ell \, || \nabla \vect{B} ||^2, \\
\label{eq:f_grad_grad}
f_{\nabla\nabla} &=& \frac{1}{L} \int d\ell \, || \nabla \nabla \vect{B} ||^2, \\
\label{eq:f_B2}
f_{B2} &=& \frac{1}{L} \int d\ell \left[ B_{20} -  \left(\frac{1}{L} \int d\ell \; B_{20} \right) \right]^2 \\
f_{\mathrm{well}} &=& \max\left( 0, \; \frac{d^2 V}{d\psi^2} - W_* \right)^2, \\
f_{\mathrm{Merc}} &=& \max\left( 0, \; D_* - \DMerc \right)^2.
\end{eqnarray}
Here, $L$ is the length of the magnetic axis,  $\int d\ell$ indicates an integral over the axis, and quantities with a subscript $*$ indicate specified target values.
The motivation for these terms is as follows.

Minimizing $f_{\nabla}$ increases $L_{\nabla B}$, 
\changed{which in practice is found to expand}
the radius of good quasisymmetry. The term $f_{\nabla}$ is the most effective term for this purpose based on experience so far.
Similarly, minimizing $f_{\nabla\nabla}$ increases the radius of good quasisymmetry by increasing 
$L_{\nabla \nabla B}$.
Note that the term $f_{\nabla}$ does not depend on the second-order solution at all, so it does not directly constrain $B_{2c}$. 

The term $f_L$ is included to avoid a problem that otherwise occurs when the initial axis shape is consistent with quasi-helical symmetry. (I.e., the normal vector makes complete poloidal rotations as the axis is followed toroidally). In this case, the optimizer can reduce $f_\nabla$ by making the helical excursion of the magnetic axis as large as the major radius, so $R$ drops to 0 every field period. This state is unacceptable, since adequate space is required in the middle of the torus for the electromagnetic coils and other components. By including $f_L$ in the objective, this problem is avoided. An objective term that penalizes values of $R$ below a threshold was also considered to avoid this pathology. However $f_L$ has produced better optima in practice. Through different choices of $L_*$, the user can parameterize a family of configurations in which there is a trade-off between the quality of quasisymmetry versus space in the middle of the torus.

Similarly, the term $f_\iota$ is included to avoid a problem that otherwise occurs when the initial axis shape is consistent with quasi-axisymmetry. (I.e., the normal vector does not make any complete poloidal rotations as the axis is followed toroidally). In this case, the optimizer can reduce $f_\nabla$ by making the axis axisymmetric. Including $f_\iota$ in the objective with a nonzero value of $\iota_*$ cures this problem. Because $f_L$ and $f_\iota$ are useful for QH and QA symmetry respectively, we set $w_L=0$ when seeking QA configurations and set $w_\iota=0$ when seeking QH configurations.

Minimizing the term $f_{B2}$ makes $B_{20}$ (nearly) independent of $\varphi$. This makes the near-axis solution fully quasisymmetric through second order in $r$.

\begin{figure*}
  \centering
\includegraphics[width=\columnwidth]{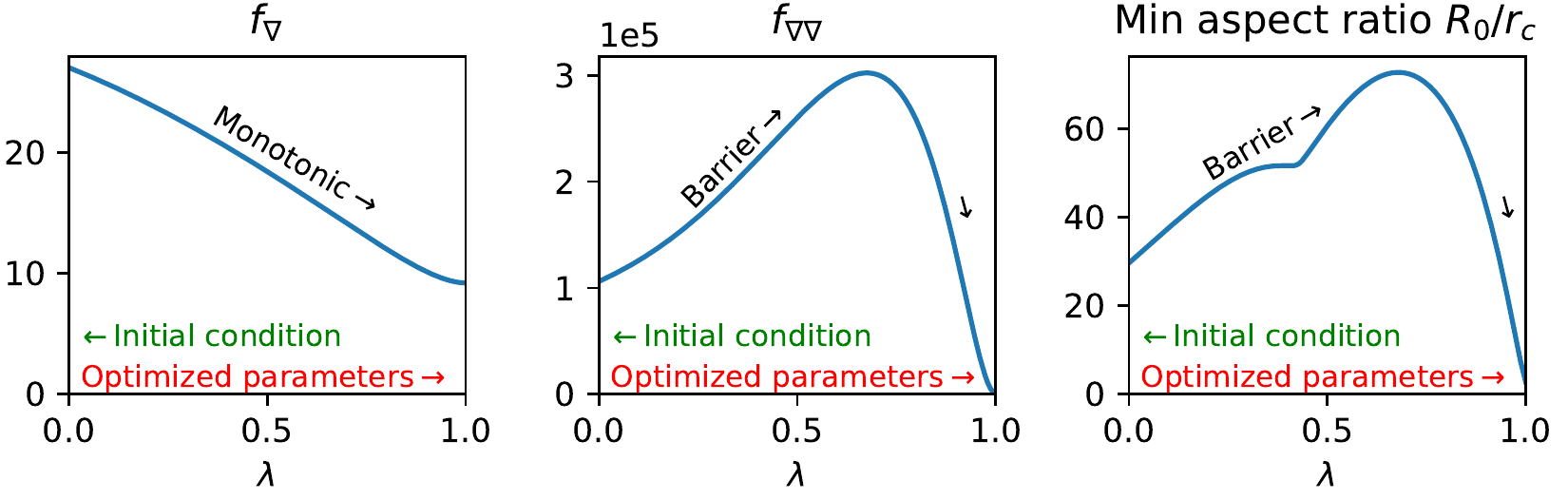}
  \caption{\label{fig:barrier} 
Variation of several possible objective function terms as one interpolates between the optimized QH configuration of \cite{r2GarrenBoozer} ($\lambda=1$)
and a typical initial condition ($\lambda=0$) defined by $R_n=Z_n=0$ for $n>1$, $\etabar=1$, and $B_{2c}=0$.
The objective $f_\nabla$ is most useful because it varies between these configurations monotonically. The other terms shown increase with $\lambda$ before they decrease, making it hard to get from $\lambda=0$ to $\lambda=1$ when they dominate the objective.
}
\end{figure*}

Although we in principle wish to maximize $r_c$, the minor radius at which the second-order surfaces become singular, we find it not very effective in practice to directly optimize functions of $r_c$.
A possible reason for this can be understood from figure \ref{fig:barrier}. The horizontal coordinate $\lambda$ in this figure indicates an interpolation between the optimized QH configuration of section 5.4 of \cite{r2GarrenBoozer}, corresponding to $\lambda=1$, and a typical initial condition, corresponding to $\lambda=0$. Letting a subscript * denote values for the optimized configuration, the parameters for the intermediate configurations are $R_n = \lambda R_{n*}$ and $Z_n=\lambda Z_{n*}$ for $n>1$, $R_n=R_{n*}$ and $Z_n=Z_{n*}$ for $n \le 1$, $\etabar = 1 + \lambda (\etabar_* - 1)$, and $B_{2c}=\lambda B_{2c*}$. Therefore, as $\lambda$ decreases from 1 to 0, the configuration is smoothly interpolated to one with simplified parameters, such as only Fourier modes with $n =0$ or 1 in the axis shape. In optimization, one typically moves in the opposite direction, starting with an initial guess like the $\lambda=0$ case, and aiming to end up at a configuration like the $\lambda=1$ case. It can be seen in figure \ref{fig:barrier} that $R_0/r_c$ is not monotonic along this path. Due to the ``barrier'' in between, it is hard to get from the $\lambda=0$ initial condition to the $\lambda=1$ optimized configuration using the objective $R_0/r_c$, even though the final value of $R_0/r_c$ at $\lambda=1$ is favorable. In contrast, the figure shows that $f_\nabla$ is monotonically decreasing with $\lambda$, making it an effective objective function.
In short, 
although minimizing the aspect ratio $R_0/r_c$ is an intuitive goal, it turns out that directly applying minimization to $R_0/r_c$ is less effective than minimizing the better-behaved 
function $f_\nabla$.

Figure \ref{fig:barrier} also shows
the function $f_{\nabla\nabla}$ has a similar non-monotonic behavior to $R_0/r_c$, so $f_{\nabla\nabla}$ is ineffective if used as the dominant term in the objective function. However adding a small multiple of $f_{\nabla\nabla}$ or $R_0/r_c$ to $f_\nabla$ can still be effective. Other potential objective functions were also explored based on $X_2$, $Y_2$, and their $d/d\varphi$ derivatives. These quantities were found to have non-monotonic behavior similar to the right two panels of figure \ref{fig:barrier}. 
\changed{
A possible 
explanation for the different monotonic vs non-monotonic behavior of the various objective function terms might be the following.
Both $f_{\nabla\nabla}$ and
$R_0/r_c$ depend on $O(r^2)$ quantities whereas $f_{\nabla}$ depends only on $O(r^1)$ quantities, and the $O(r^2)$ quantities are more sensitive to small changes in the axis shape due to $d/d\varphi$ derivatives in the near-axis equations.
}
In practice, the most effective approach to lower the aspect ratio seems to be using an objective dominated by $f_\nabla$, with a small multiple of $f_{\nabla\nabla}$ added after an initial minimization of $f_{\nabla}$.

The term $f_{\mathrm{well}}$ can be included to obtain configurations with a vacuum magnetic well, $d^2 V / d \psi^2 < 0$. Typically $W_*$ is set to a negative value to provide some margin against instability. Similarly, for configurations with pressure and/or current, $f_{\mathrm{Merc}}$ can be included to obtain Mercier-stable configurations. The value $D_*$ should be set to a positive value to provide a stability margin.
It is unclear whether magnetic well and/or Mercier stability should be included in stellarator design, since multiple experiments have reported operating in unstable regimes without major difficulty
\citep{Geiger,Watanabe,Weller2006,Aguilera}.

\changed{
Whether an optimization produces a QA or QH configuration is determined by the initial condition for the axis shape.
In a QH configuration with given $N$, the axis normal vector rotates poloidally about the axis $N$ times as one traverses the axis toroidally (as discussed in section 5.2 of \cite{PaperI}). QA configurations represent the $N=0$ case: the normal vector makes no net rotations about the axis as the axis is traversed toroidally.
If the axis is continuously deformed from one $N$ value to another, the curvature crosses through zero, causing $f_\nabla$, $f_{\nabla\nabla}$, and $f_{B2}$ to diverge. This results in an infinitely steep barrier in the objective function, which the optimizer will not cross. Hence the symmetry class ($N$ value) of the optimum will match that of the initial condition.
}

The integrals (\ref{eq:f_grad})-(\ref{eq:f_B2}) are discretized using a uniform grid in the standard toroidal angle $\phi$ with $N_\phi$ points.
Upon discretization, all the terms in $f$ have the form of a sum of squares. 
This is true also for the terms that are integrals over the axis,
for instance,
\begin{equation}
\label{eq:f_grad_discrete}
f_{\nabla} =  \sum_{i=1}^{N_\phi}\sum_{j=1}^3 \sum_{k=1}^3 
\left[ \sqrt{\frac{\Delta\phi}{L}\frac{d\ell}{d\phi}} (\nabla\vect{B})_{j,k} \right]^2.
\end{equation}
Here, $j$ and $k$ range over the $\vect{t}$, $\vect{n}$, $\vect{b}$ components, $\Delta\phi$ is the grid spacing in $\phi$, and the quantity in large square brackets is evaluated at the toroidal grid point $i$.
The other terms in the objective involving integrals are discretized as 
\begin{equation}
f_{\nabla\nabla} =  \sum_{i=1}^{N_\phi}\sum_{j=1}^3 \sum_{k=1}^3 \sum_{n=1}^3 
\left[ \sqrt{\frac{\Delta\phi}{L}\frac{d\ell}{d\phi}} (\nabla\nabla\vect{B})_{j,k,n} \right]^2
\end{equation}
and
\begin{equation}
\label{eq:f_B2_discrete}
f_{B2} =  \sum_{i=1}^{N_\phi}
\left[ \sqrt{\frac{\Delta\phi}{L}\frac{d\ell}{d\phi}} \left(B_{20} - \bar{B}_{20}\right) \right]^2.
\end{equation}
Therefore the discretized problem can be solved using methods for nonlinear least-squares problems.
The quantities in square brackets in (\ref{eq:f_grad_discrete})-(\ref{eq:f_B2_discrete}) are the residuals for the least-squares problem.

It is effective to \changed{increase the dimensionality of} the parameter space in several steps. 
For the first step, a maximum mode number $N_F=1$ is used, with $N_F$ incremented by one each step. For $N_F = 1,2,3$, the weights $w_{\nabla\nabla}$, $w_{\mathrm{well}}$, and $w_{\mathrm{Merc}}$ are set to zero, which is found to make the optimization very robust. These weights are set to nonzero values if desired for later steps. As the number of Fourier modes in the parameter space is increased, the number of grid points $N_\phi$ can be increased as well (as is done for results here).

We solve the optimization problem using the C++ implementation at
\url{https://github.com/landreman/qsc}, also archived at \cite{zenodoData}. Results here are obtained with the Levenberg-Marquardt algorithm implemented in the GNU scientific library \citep{GSL}.


\section{Parameter scans}
\label{sec:scan}

Parameter scans are applied to the optimization problem of section \ref{sec:opt}, to understand the set of possible quasisymmetric configurations. 
The parameters scanned include $\nfp$, the $w_j$ weights in eq (\ref{eq:objective}), the target values $\iota_*$, $L_*$, $W_*$, and $D_*$, the pressure $p_2$, whether or not magnetic well or Mercier stability is imposed, and whether QA or QH is sought.
The number of field periods is scanned from one through 10. 
In each case, the initial axis shape before optimization is $R(\phi)=1+\Delta \cos(\nfp \phi)$ and $Z(\phi)=\Delta \sin(\nfp \phi)$ for a chosen number $\Delta$.
When searching for QA solutions, $\Delta$ is chosen $<1/(\nfp^2+1)$ so the normal vector does not make complete poloidal rotations, $w_L$ is set to 0, and $\iota_*$ is scanned. When searching for QH solutions, $\Delta$ is chosen $>1/(\nfp^2+1)$ so the normal vector makes $\nfp$ complete poloidal rotations, $w_\iota$ is set to 0, and $L_*$ is scanned.
These conditions on $\Delta$ have been discussed recently by \cite{RodriguezPhases}.
The weight $w_\nabla$ is always 1, since  $f_\nabla$  is the most reliable term to include in the objective, as discussed in the previous section. The other weights are scanned logarithmically over several orders of magnitude.

For many choices of weights and targets, the configuration at the end of an optimization may be unacceptable if the volume of good quasisymmetry is too small, the rotational transform is too low, the elongation is too large, etc. Therefore as the parameters are scanned, configurations are saved only if they pass through several filters, i.e. satisfy several inequalities. 
One such filter is $|\iota| > 0.2$; sufficient $\iota$ is required since the equilibrium $\beta$ limit scales $\propto \iota^2$, and the width of banana orbits in QA scales $\propto 1 / \iota$.
Other typical inequalities imposed are $L_{\nabla B} > 0.2 R_0$, $L_{\nabla\nabla B} > 0.2 R_0$, the variation of $ B_{20}$ is $< B_0$, elongation in the plane perpendicular to the axis $< 10$ (computed from the $O(r^1)$ elliptical surfaces), and minimum minor radius $r_c > 0.15 R_0$. 
A minimum $R(\phi)/R_0$ is enforced, e.g. $> 0.4$, to ensure some space for coils near the coordinate origin.
The quantities 
$|X_{20}|$, $|X_{2s}|$, $|X_{2c}|$, 
$|Y_{20}|$, $|Y_{2s}|$, $|Y_{2c}|$, 
$|Z_{20}|$, $|Z_{2s}|$, and $|Z_{2c}|$ are required to be below a threshold such as 10.0. 
This constraint is another heuristic method for ensuring the radius of applicability of the asymptotic series is relatively large, by ensuring the $O(r^2)$ terms are not too much larger than the $O(r^1)$ terms.
Similarly, the quantities 
$|\partial X_{20} / \partial\varphi|$,
$|\partial X_{2s} / \partial\varphi|$,
$|\partial X_{2c} / \partial\varphi|$,
$|\partial Y_{20} / \partial\varphi|$,
$|\partial Y_{2s} / \partial\varphi|$,
$|\partial Y_{2c} / \partial\varphi|$,
$|\partial Z_{20} / \partial\varphi|$,
$|\partial Z_{2s} / \partial\varphi|$, and
$|\partial Z_{2c} / \partial\varphi|$
are required to be below a threshold such as 20.0. 
The exact values of the thresholds are adjusted from case to case; for instance it is harder to find high-$\beta$ configurations with Mercier stability so generous thresholds are used in this case. In contrast, vacuum configurations without a magnetic well constraint are comparatively easier to obtain, so more restrictive thresholds are used in this case to focus on the most interesting solutions.

\begin{figure*}
  \centering
\includegraphics[width=\columnwidth]{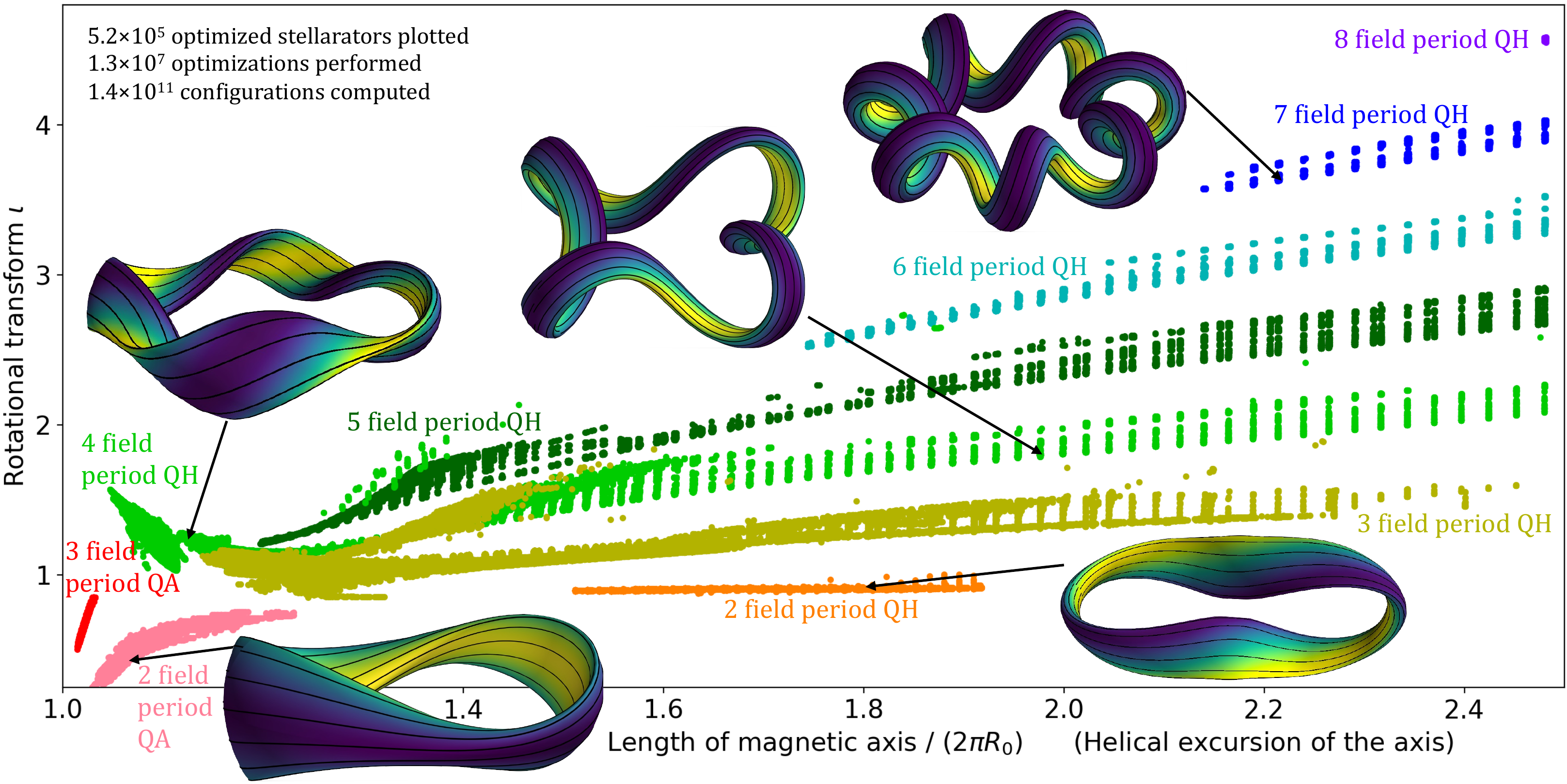}
  \caption{\label{fig:landscape} 
Results from optimizing the near-axis parameters, scanning the weights, targets, and $\nfp$. Each point indicates the result of an optimization. Five representative points are highlighted with visualizations of the flux surfaces in 3D and are discussed in section \ref{sec:configs}.
}
\end{figure*}

The results of the parameter scans are shown in figure \ref{fig:landscape}.
Each point indicates an independent optimization for specific choices of weights and target values. The points are colored to indicate $\nfp$ and  QA vs QH symmetry.
The horizontal coordinate is the length of the magnetic axis, normalized so that 1 indicates a circle, while larger values indicate greater helical excursion of the axis. 
The axes of the figure were taken to be axis length vs $\iota$ since this choice effectively separates the data into clusters. 
The apparent stripes at fixed axis length are an artifact of the grid of target $L_*$ values in the scans.
The database of configurations includes both vacuum and finite-beta cases (i.e. various choices of $p_2$), with and without magnetic well, and with and without Mercier stability.

A total of $5\times 10^5$ points are plotted in figure \ref{fig:landscape}.
As discussed above, most optimizations resulted in configurations that were filtered out, so a total of $> 10^7$ optimizations were run to produce the figure.
Each optimization involved many evaluations of the objective function, eq (\ref{eq:objective}), so a total of $> 10^{11}$ evaluations of the objective were performed to produce the figure. 
Each function evaluation typically takes under 1 ms.
The speed by which the near-axis equations can be evaluated makes it possible to evaluate this very large number of configurations in tens of wallclock hours on a computing cluster.

Many interesting patterns can be seen in the data.
QA solutions are limited to a single continuous band for each $\nfp$ at the lower left, with $\iota < 1$ and relatively circular magnetic axis. QH solutions are found at a wide range of $\iota$, from just below 1 up to $>4$. The QH solutions for any given $\nfp$ also occupy continuous bands with a wide range of axis lengths. Analogous features have been observed recently by \cite{RodriguezLandscape}.

As with QH symmetry, QA symmetry scans were performed for all values of $\nfp$ from 1 through 10. However no configurations passed the filters for $\nfp > 3$. 
\changed{(For example, for QA with $\nfp=4$ and no constraints other than $\iota>0.2$, the largest $r_c$ obtained was only $0.11 R_0$.)}
These findings are consistent with previous reports of QA only for $\nfp=2$ and 3. 
Furthermore, in the scans here, the QA configurations for $\nfp=3$ had smaller values of $r_c$ than the $\nfp=2$ QAs: specifically $r_c$ was $<0.2$ for $\nfp=3$, whereas $r_c$ attained values up to 1.0 for $\nfp=2$. Therefore in the $\nfp=3$ configurations, the region of good quasisymmetry is limited to a higher aspect ratio.
This finding is consistent with the fact that previous optimizations for quasi-axisymmetry at $\nfp=3$ (NCSX and ARIES-CS) have had significant imperfections in the symmetry, whereas excellent quasi-axisymmetry has been obtained at $\nfp=2$ \citep{LandremanPaul2022,GiulianiSurface}. When $\nfp=1$, QA solutions were found that satisfied all constraints, but they resembled $\nfp=2$ configurations that were translated or rotated to break two-field-period symmetry. These configurations did not appear to have an advantage over $\nfp=2$ configurations and so will not be considered further.

In contrast, QH solutions were found that passed the filters for all values of $\nfp$ attempted except 1. For QH solutions, the number of field periods for which the axis length can be minimized is 4, followed closely by 3 and 5. For other values of $\nfp$, QH solutions require significant helical excursion of the axis and an associated longer axis length.

\changed{
A related but different parameter scan was shown previously in figure 2 of \citep{Boozer2020}. That previous scan used only the $O(r)$ terms in the expansion rather than $O(r^2)$, and the Fourier modes of the axis were scanned directly, with no optimization applied. The filters used in the present figure \ref{fig:landscape} eliminate significant parts of the parameter space from the earlier scan. This can be seen for example in the more limited ranges of $\iota$ for each value of $N$ in the present figure \ref{fig:landscape} compared to the previous scan.
}

For five of the points in figure \ref{fig:landscape}, the flux surface shapes of the associated optimized configurations are shown in the same figure in 3D. These configurations and others from the scan are discussed in greater detail in section \ref{sec:configs}. Of these highlighted configurations, the two on the left are relatively familiar in shape: a two-field-period QA resembling CFQS \citep{CFQS}, and a four-field-period QH resembling HSX \citep{HSX}.
One of the other highlighted configurations is a four-field-period QH with large helical excursion of the axis, more excursion than in previously described QH stellarators aside from the recent configuration by \cite{RodriguezLandscape}.
The other two configurations shown in 3D have shapes unlike previously reported quasisymmetric stellarators. These include QH configurations with unusual numbers of field periods, two and seven.


\section{Example configurations}
\label{sec:configs}

We now present several specific configurations obtained using the parameter scans in section \ref{sec:scan}. 
All input and output files for these configurations and the optimizations that led to them can be found in the supplemental material on Zenodo \citep{zenodoData}.

For each configuration, a finite 
aspect ratio is chosen for the figures. The finite aspect ratio boundary is generated as described in section 4.2 of \cite{PaperII}. Namely, a finite value $a$ is chosen for the minor radius variable $r$, the position vector (\ref{eq:positionVector}) is evaluated for $r=a$, and the result is converted to Fourier series in cylindrical coordinates. 
\changed{To evaluate the position vector, some $O(r^3)$ terms are included, as detailed in section 3 of \cite{r2GarrenBoozer}.}
For each boundary surface, several definitions of the aspect ratio are available. In the near-axis equations, a convenient definition of aspect ratio is $R_0/a$. However this definition differs from the widely used definition of aspect ratio in the stellarator community, $A=\bar{R}/\bar{a}$, which is obtained as follows. The effective minor radius $\bar{a}$ is defined by setting the toroidally averaged cross-sectional area of the shaped boundary equal to the area of a circle with minor radius $\bar{a}$. Then the effective major radius $\bar{R}$ is defined by setting the volume of the shaped boundary equal to that of a circular cross-section axisymmetric torus with major radius $\bar{R}$. (See page 12 of \cite{r2GarrenBoozer} for details.)

To confirm the correctness of the near-axis method, each example below is checked using a fixed-boundary MHD equilibrium calculation that does not make a near-axis expansion, as follows.
Given the constructed boundary, the field inside is computed and converted to Boozer coordinates with the DESC code \citep{Desc0, Desc1, Desc2, Desc3}.
Similar checks of near-axis solutions were done using the VMEC code \citep{VMEC1983} previously in \cite{PaperII, PaperIII, r2GarrenBoozer,LandremanMercier}.
It was shown in this previous work that $\iota$, $d^2V/d\psi^2$, $D_{\mathrm{Merc}}$, and Fourier modes of $B$ for fully 3D equilibria converged to the values predicted by the near-axis solution as the aspect ratio increased.
No further optimization is applied to the finite-aspect-ratio configurations here, although this could be done in future work.

\changed{
For the examples that follow, the finite minor radius $a$ is chosen by hand based on several considerations. The spectral width of the constructed boundary should be sufficiently small that the DESC calculations reach acceptable force residuals with poloidal and toroidal mode numbers $\le  12$. This is easier to achieve for smaller $a$. Also $a$ is chosen to be sufficiently small that $B$ from the DESC solution is reasonably similar to the near-axis prediction.
}


\subsection{Quasi-axisymmetry with two periods}
\label{sec:QA}



\begin{figure*}
  \centering
\includegraphics[height=2.5in]{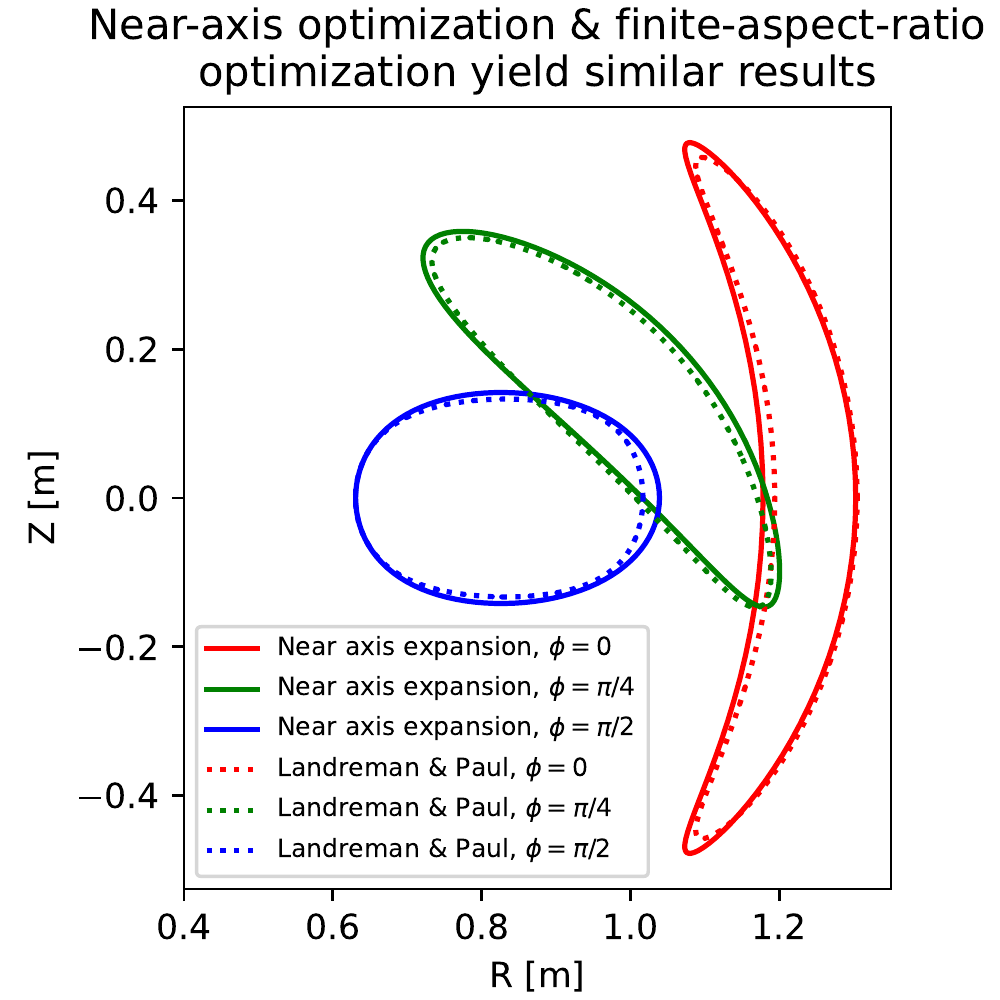}
\includegraphics[height=2.5in]{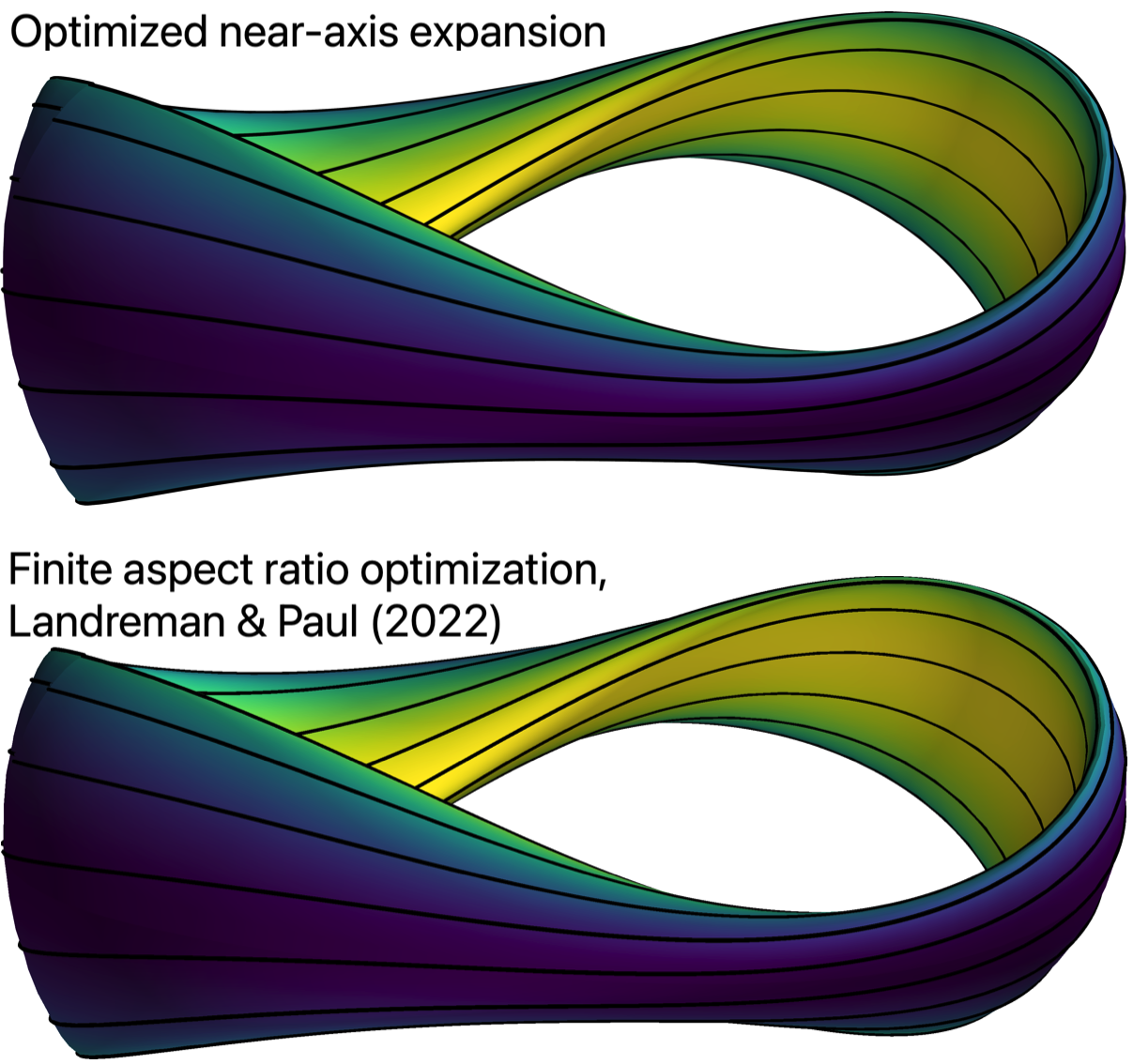}
  \caption{\label{fig:QA_xsections} 
Optimization of near-axis solutions and  traditional finite-aspect-ratio optimization, run completely independently, can yield similar results. Here both methods are used to generate a two-field-period vacuum QA with $\iota\approx 0.42$ and magnetic well at aspect ratio 6.0. Left: Cross sections of the results are plotted at three toroidal angles, showing the similar surface shapes. 
Right: Views of the same configurations in 3D.
The finite-aspect-ratio optimization is from \cite{LandremanPaul2022}.
}
\end{figure*}

\begin{figure*}
  \centering
\includegraphics[width=\columnwidth]{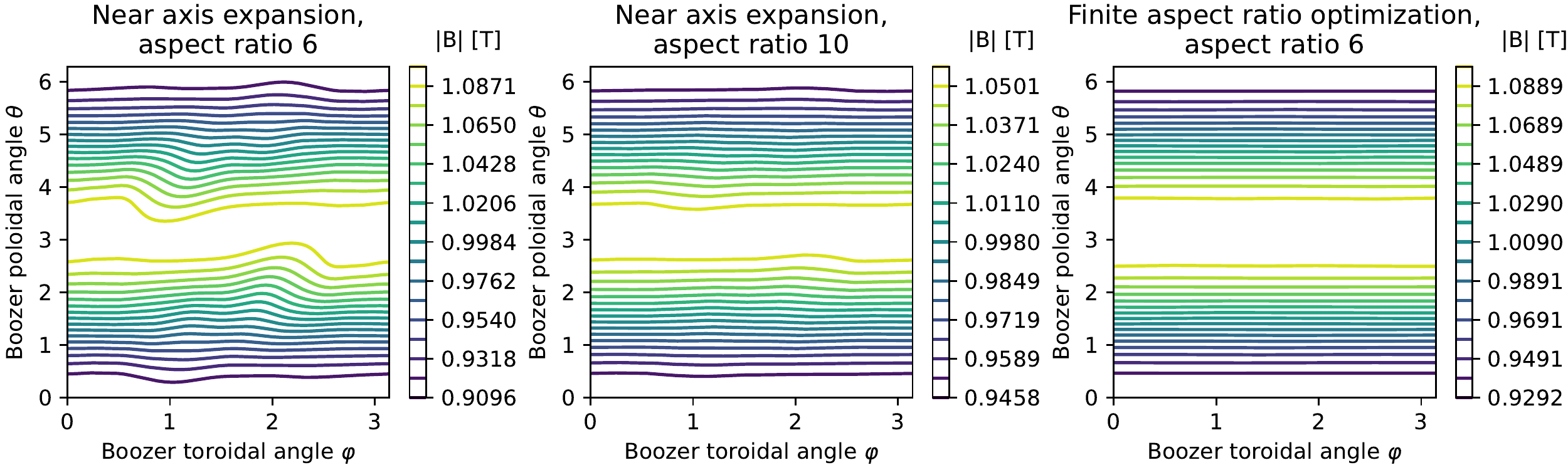}
  \caption{\label{fig:QA_Booz} 
By computing the field inside the boundary constructed from the near-axis solution, it can be confirmed that approximate QA symmetry was indeed achieved. The symmetry errors decrease as the aspect ratio increases, as expected.
For comparison, 
the finite-aspect-ratio ``QA+well'' optimization from \cite{LandremanPaul2022} is shown at right.
}
\end{figure*}

The first configuration is one that is very similar to the QA with magnetic well presented in \cite{LandremanPaul2022}.
That configuration was a two-field-period vacuum field optimized for $\iota \sim 0.42$, and so in the near-axis calculation we set $p_2=0$ and $\iota_*=0.42$.
Other than these values, the near-axis optimization is completely independent of the configuration and optimization in \cite{LandremanPaul2022}.
For the first three steps of the near-axis optimization, in which the number of Fourier modes is increased from $N_F=1$ to $N_F=3$, the only nonzero weights are $w_\nabla=1$, $w_\iota=100$, and $w_{B2}=0.01$. Then in steps with $N_F=3-7$, the weights used are $w_\nabla=w_{\nabla\nabla}=w_{\mathrm{well}}=1$, $w_{\iota}=100$, and $w_{B2}=30$. A target magnetic well of $W_*=-20$ is used to provide some margin.
\changed{(The DESC calculations at finite aspect ratio had less magnetic well than the near-axis solution, so $W_*=-20$ was found to be sufficient to achieve a magnetic well at all radii in the DESC solution.)}
For later configurations in this paper, the optimization weights and target values can be found in \cite{zenodoData}.
The full multi-stage optimization takes a total of 0.4 seconds on 1 cpu of a standard MacBook laptop. The flux surface shape at aspect ratio $A=6$ is displayed in figure \ref{fig:QA_xsections}, matching the aspect ratio used in \cite{LandremanPaul2022}. Cross sections and 3D renderings of the same two configurations, scaled to the same average major radius, are also shown in figure \ref{fig:QA_xsections}. It can be seen that the surface shape generated by the near-axis method is qualitatively similar to the one obtained independently by finite aspect ratio optimization. 
The field strength computed by DESC on the aspect ratio 6 boundary is displayed as a function of the Boozer angles in figure \ref{fig:QA_Booz}. It can be seen that QA is achieved approximately, though not as accurately as it is with finite aspect ratio optimization. The figure also shows a similar calculation for the boundary constructed at a higher aspect ratio, 10, showing that the symmetry improves at higher aspect ratio as expected. Indeed, as the aspect ratio is increased, QA can be achieved to any desired precision, as demonstrated in \cite{r2GarrenBoozer}. Overall, we can conclude that while the near axis approach is not as accurate as finite aspect ratio optimization, it can compute qualitatively similar configurations extremely fast.

\subsection{Quasi-helical symmetry with two periods}

One noteworthy discovery from the parameter scan is that there are QH solutions with only two field periods. To our knowledge, two field period QH configurations have not been reported previously. These configurations may be attractive since the number of modular coils tends to scale with the number of field periods. Therefore a two-field-period QH may require fewer coils than other QH configurations, reducing cost and enabling greater access between coils. At the same time, QH configurations can have very good confinement of energetic particles
\citep{Bader2021,LandremanPaul2022,Paul2022} due to the thinner banana orbits and related factors.

\begin{figure*}
  \centering
\includegraphics[width=2.0in]{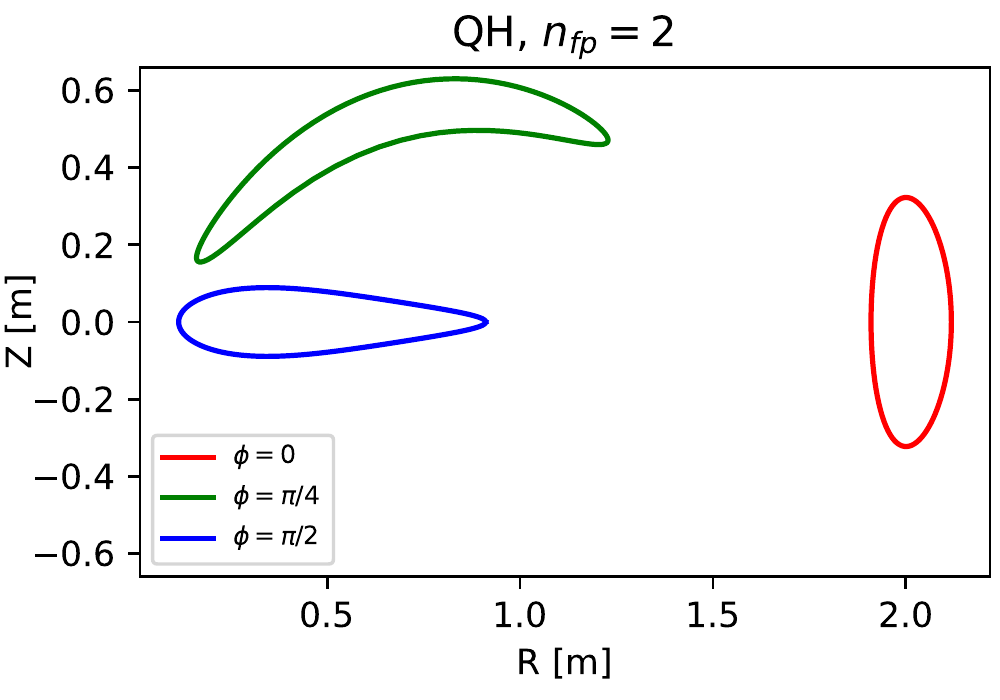}
\hspace{0.1in}
\includegraphics[width=3.1in]{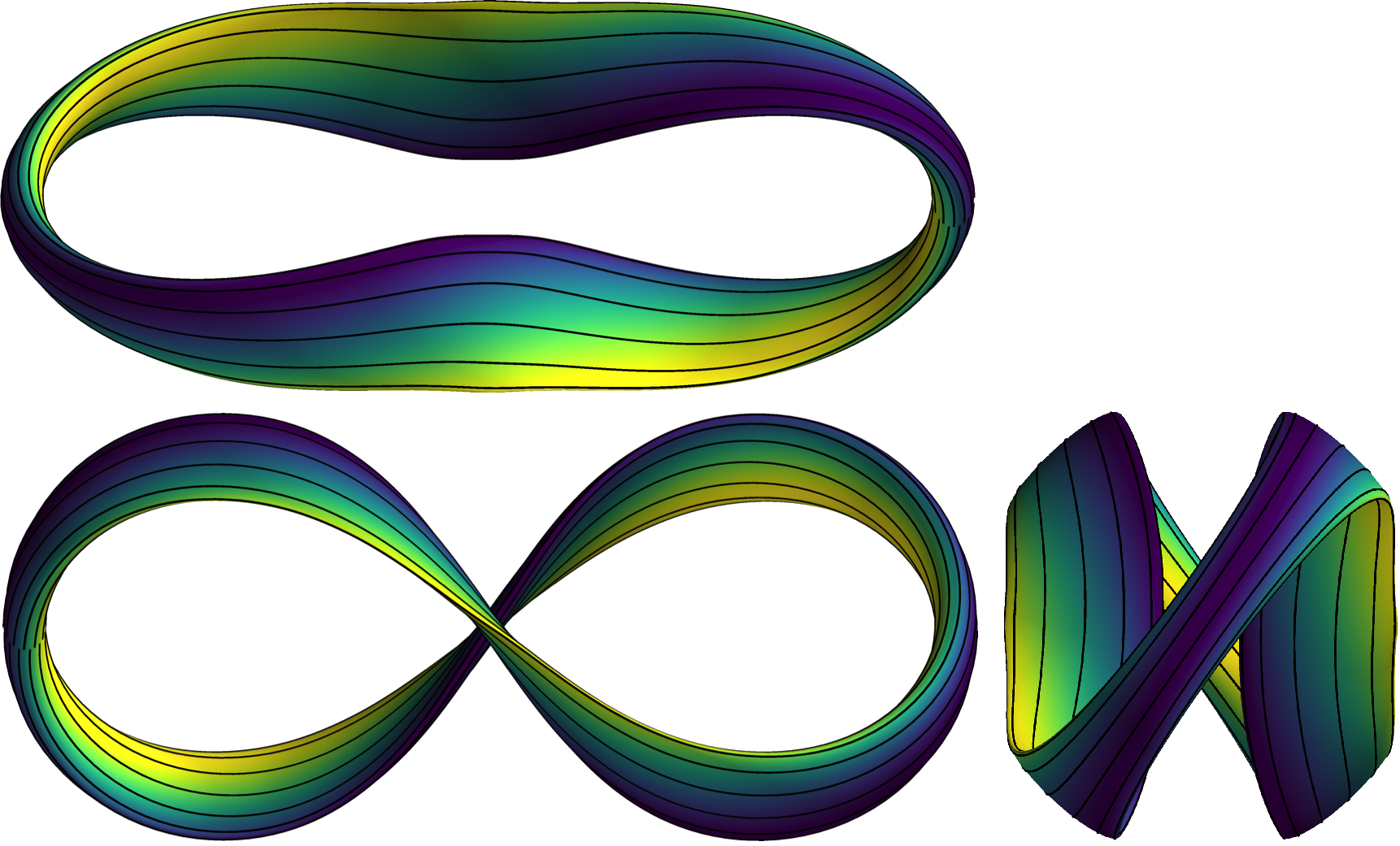}   \caption{\label{fig:nfp2_QH} 
A two-field-period quasi-helically symmetric stellarator generated from the near-axis method. Left: cross-sections. Right: The same configuration is shown from three perspectives. Color indicates the field strength, and black curves are field lines.
}
\end{figure*}

A two-field-period QH configuration generated by the optimization procedure here is shown in figure \ref{fig:nfp2_QH}. This configuration is a vacuum field with $\iota =0.95$, and the surface plotted has $a / R_{0} = 0.12$. When viewed from one side, the configuration resembles the original figure-eight design proposed by \cite{Spitzer58}. However, in contrast to Spitzer's design, the new configuration here has a non-circular cross-section yielding QH symmetry, providing improved confinement. A challenge for this new configuration is that there is not much space in the middle for coils. An important question for future research is whether $\nfp=2$ QH configurations can be found with more space in the middle, and whether feasible coil solutions exist.

It takes many toroidal Fourier modes to represent this configuration in cylindrical coordinates due to the strong shaping, with regions at small major radius and with high inclination with respect to the $z=0$ plane. It may be for this reason that $\nfp=2$ QH configurations have not been reported previously. 
Here, for the near-axis calculations, 11 Fourier modes are used to represent $R(\phi)$ and $Z(\phi)$.

\begin{figure*}
  \centering
\includegraphics[height=2.0in]{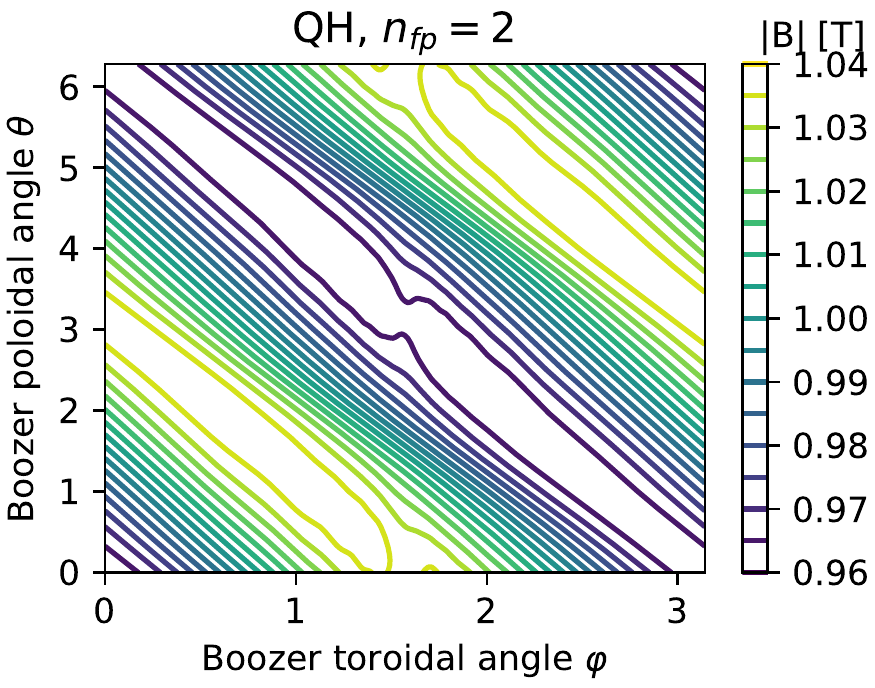}
  \caption{\label{fig:QH_nfp2_Boozer} 
  Magnetic field strength on the $a/R_0=0.05$ surface of the 
 two-field-period quasi-helically symmetric stellarator, computed by running a fully 3D fixed-boundary equilibrium calculation
inside the boundary constructed from the near-axis method.
The mostly straight diagonal contours confirm the QH symmetry. 
}
\end{figure*}

Figure \ref{fig:QH_nfp2_Boozer} shows a fully 3D calculation of $B$ for this configuration with $a/R_0=0.05$. 
It can be seen 
that the $B$ contours are (mostly) straight and diagonal, confirming the desired QH symmetry.

\subsection{Quasi-helical symmetry with three field periods}

Next we consider QH configurations in which the number of field periods is three. Previously, a configuration with these properties, obtained using conventional finite-aspect-ratio optimization, was reported in \cite{KuBoozerQHS}. Here we show two new such configurations obtained with the near-axis method.

\begin{figure*}
  \centering
\includegraphics[width=2.0in]{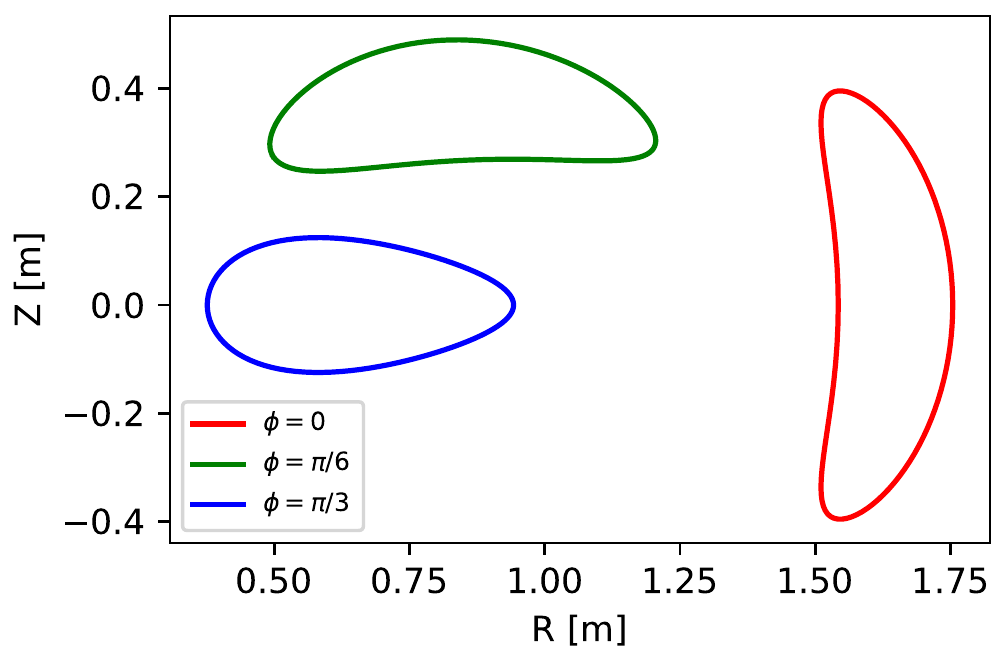}
\hspace{0.1in}
\includegraphics[width=3.0in]{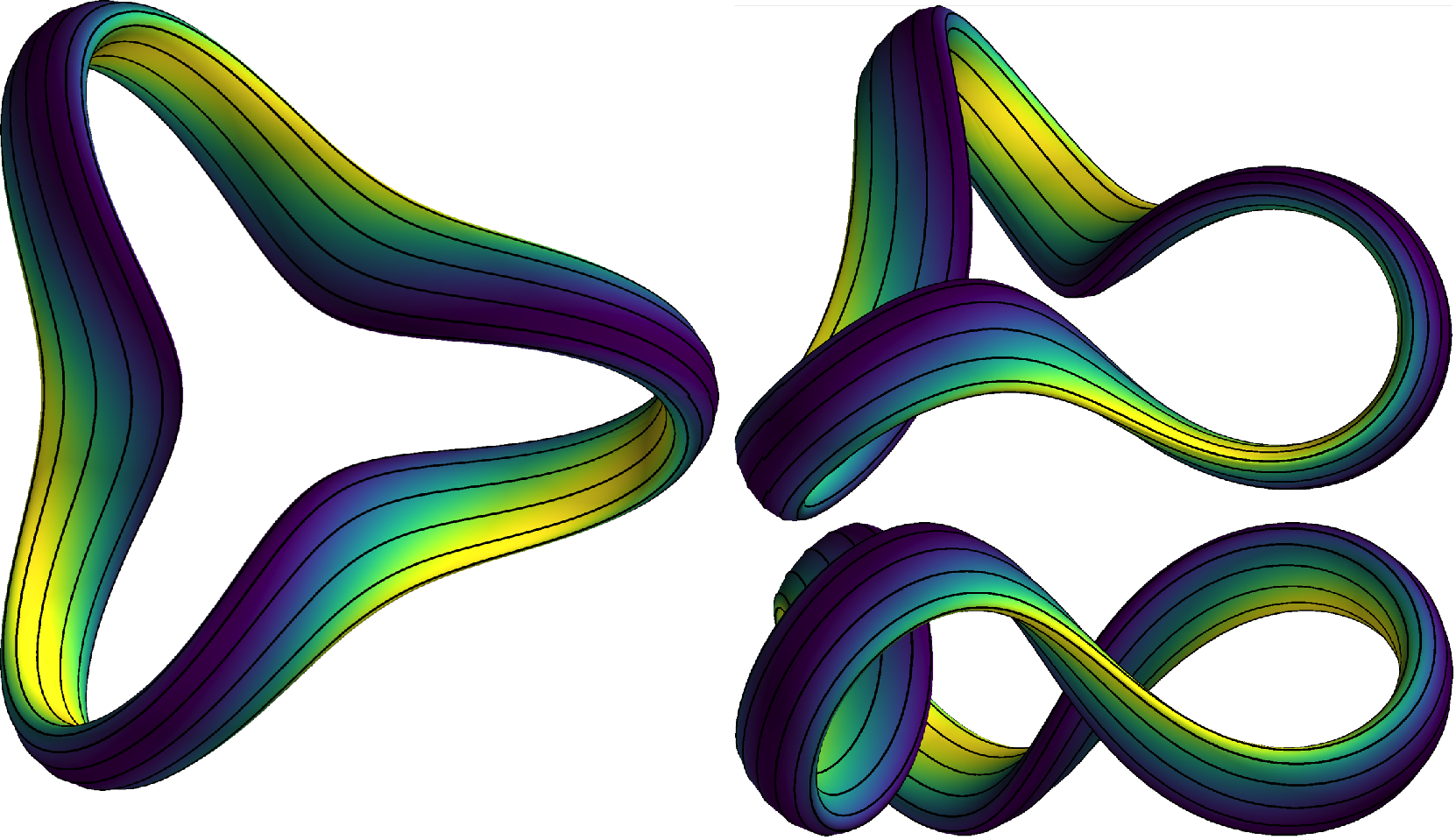}
  \caption{\label{fig:QH_nfp3_vacuum} 
A three-field-period quasi-helically symmetric stellarator generated from the near-axis method, for $\beta=0$. Left: cross-sections. Right: The same configuration is shown from three perspectives. Color indicates the field strength, and black curves are field lines.
}
\end{figure*}

\begin{figure*}
  \centering
\includegraphics[width=4.0in]{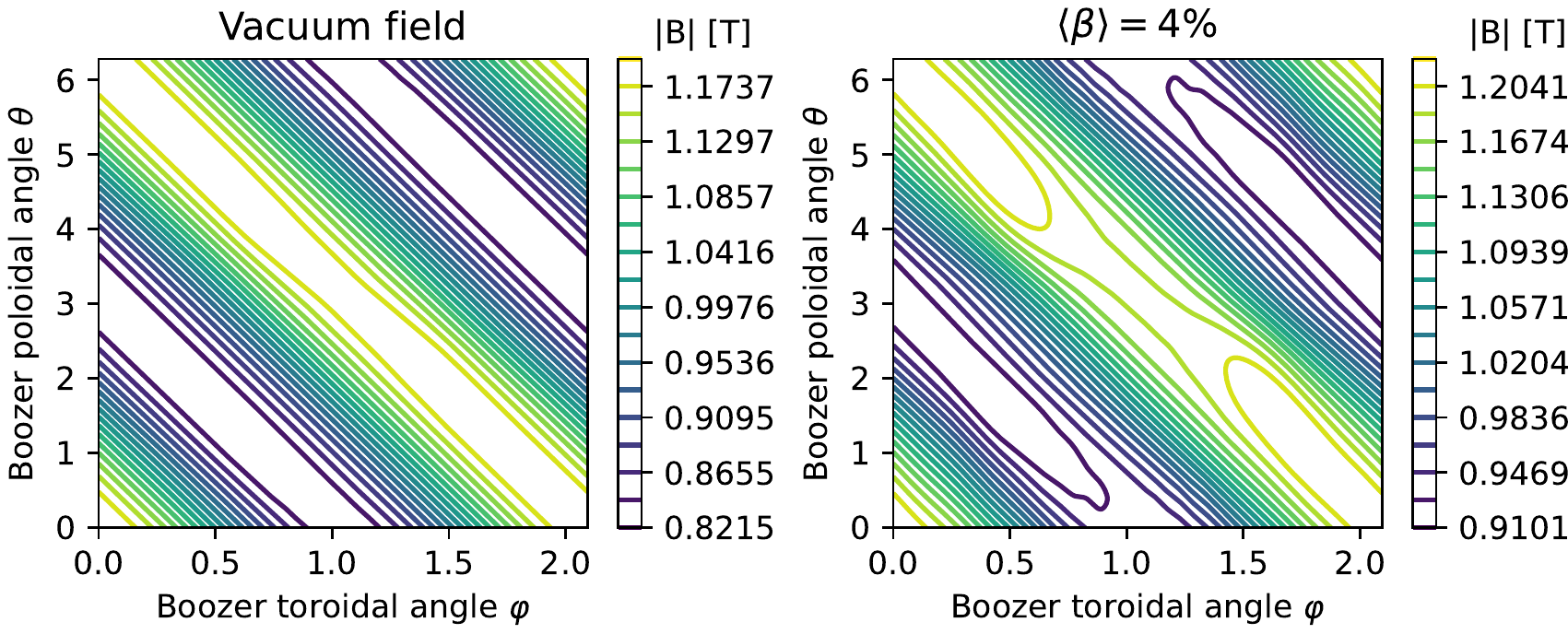}
  \caption{\label{fig:QH_nfp3_Boozer} 
Magnetic field strength on the surface of the 
 three-field-period quasi-helically symmetric stellarators, computed by running fully 3D fixed-boundary equilibrium calculations
inside the boundaries constructed from the near-axis method.
}
\end{figure*}

First, figure \ref{fig:QH_nfp3_vacuum} shows a vacuum configuration. The only terms included in the optimization were $f_L$, $f_\nabla$, $f_{\nabla\nabla}$, $f_{B2}$. 
The rotational transform obtained is $\iota = 1.25$.
For the three-dimensional views and cross-sections in figure \ref{fig:QH_nfp3_vacuum}, a minor radius of $a = 0.15 R_0$ is used, corresponding to an aspect ratio $A=5$. The field strength in Boozer coordinates on this boundary computed with DESC is shown in the left panel of figure \ref{fig:QH_nfp3_Boozer}, confirming the expected QH symmetry.

\begin{figure*}
  \centering
\includegraphics[width=2.0in]{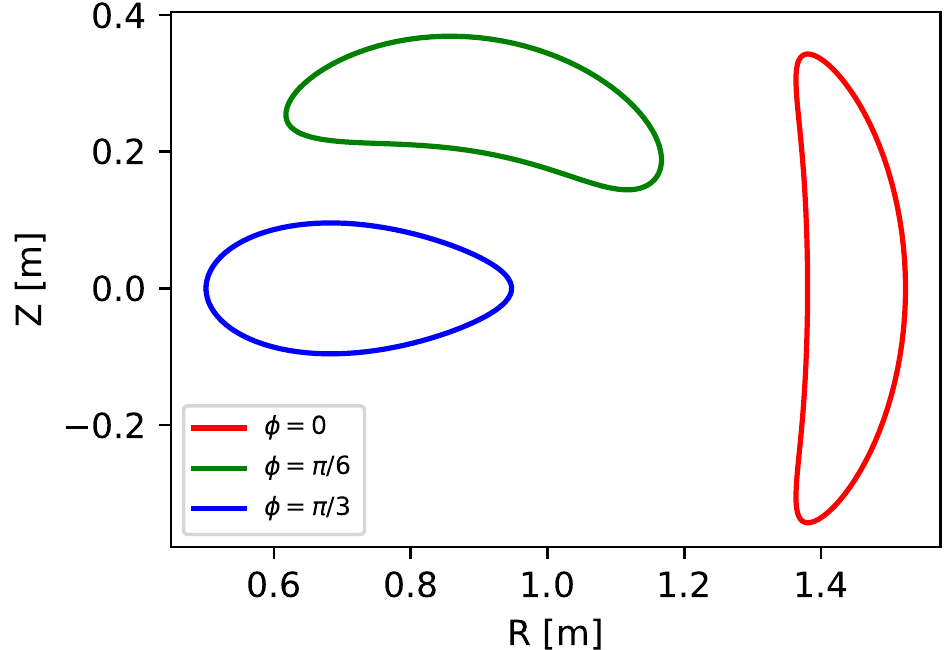}
\hspace{0.1in}
\includegraphics[width=3.0in]{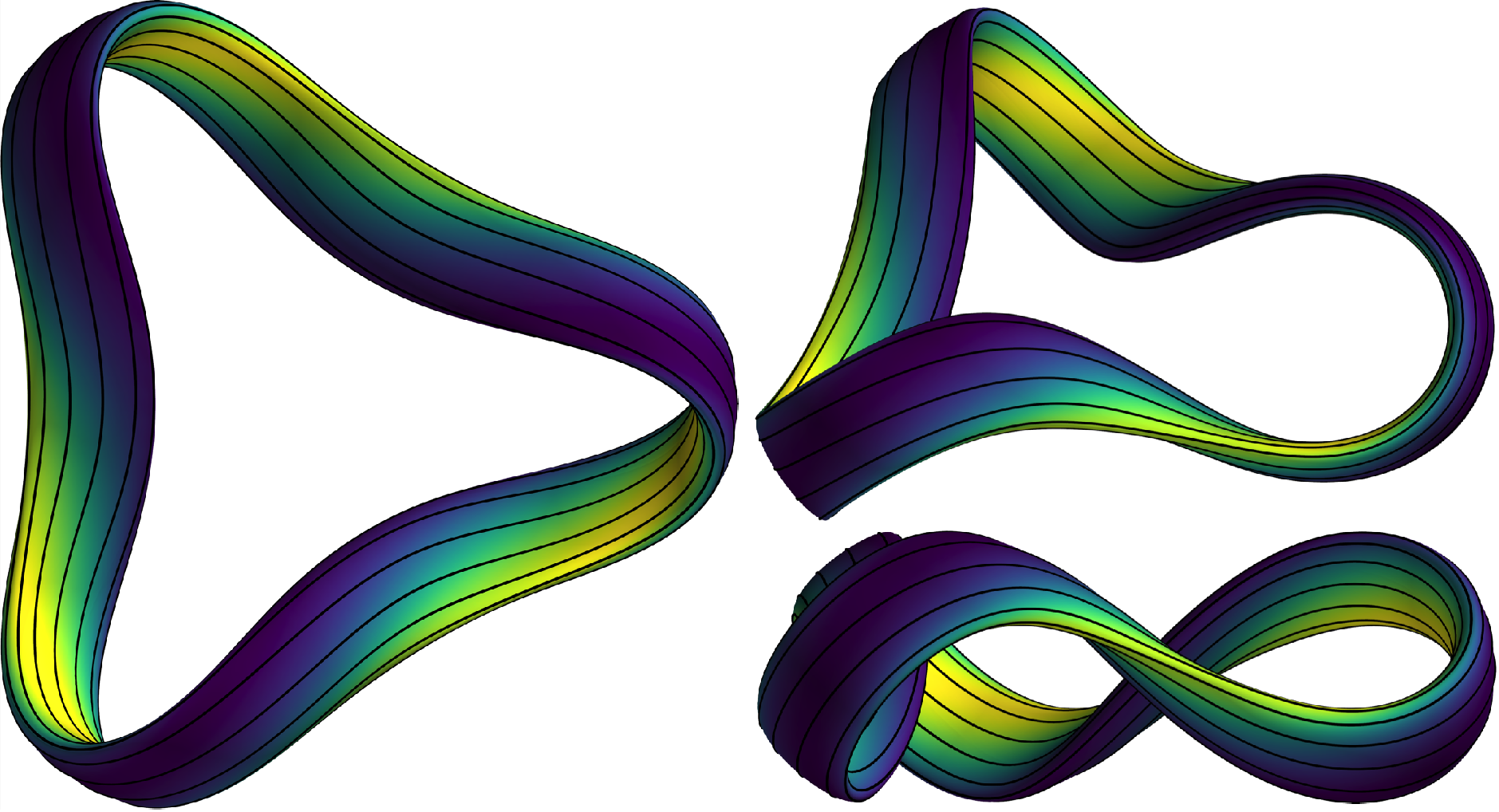}
  \caption{\label{fig:QH_nfp3_beta} 
A three-field-period quasi-helically symmetric stellarator generated from the near-axis method, for $\langle\beta\rangle=4\%$. Left: cross-sections. Right: The same configuration is shown from three perspectives. Color indicates the field strength, and black curves are field lines.
}
\end{figure*}

Second, a finite-beta configuration is shown in figure 
\ref{fig:QH_nfp3_beta}. The pressure is introduced by setting $p_2$ equal to a nonzero value, in this case $-2.0\times 10^{6}$ Pa$/$m$^2$, with the negative sign corresponding to a typical peaked profile. 
Note that for a given objective function, optimizations run at nonzero $p_2$ generally result in different axis shapes and flux surface shapes compared to optimizations with $p_2=0$.
The rotational transform for the configuration here is $\iota = 1.09$.
Note also that the absolute pressure associated with any fixed $p_2$ depends on the aspect ratio. For a pressure profile $p(r)=p_0 + r^2 p_2$ with $p=0$ at a boundary $r=a$,
the volume-averaged $\beta$ is 
$\langle\beta\rangle=2\mu_0 \langle{p}\rangle / B_0^2$ where $\langle{p}\rangle=(2/a^2)\int_0^a pr\, dr=p_0/2$ is a volume-averaged pressure, giving
$\langle\beta\rangle=-\mu_0 p_2 a^2 / B_0^2$. Hence, for a given $p_2$, a larger minor radius corresponds to a larger averaged $\beta$.
For the figures we choose a boundary minor radius $a/R_0=0.13$, slightly smaller than for the vacuum case since the QH symmetry is somewhat worse with finite pressure. At this minor radius, the aspect ratio is $A=6.5$ and $\langle \beta \rangle = 4\%$.
The field strength in Boozer coordinates on this boundary from a finite aspect ratio equilibrium calculation is shown in the right panel of figure \ref{fig:QH_nfp3_Boozer}, confirming the expected QH symmetry.
As discussed previously, a toroidal current profile $I(\psi)=0$ is used for this finite aspect ratio equilibrium calculation. 

\subsection{Quasi-helical symmetry with four field periods}

Next we consider QH configurations with four field periods. This number of field periods has been a common choice in previous QH designs \citep{HSX, KuBoozerQHS, Bader2020, LandremanPaul2022}. We will show three configurations in this category.

\begin{figure*}
  \centering
\includegraphics[width=2.0in]{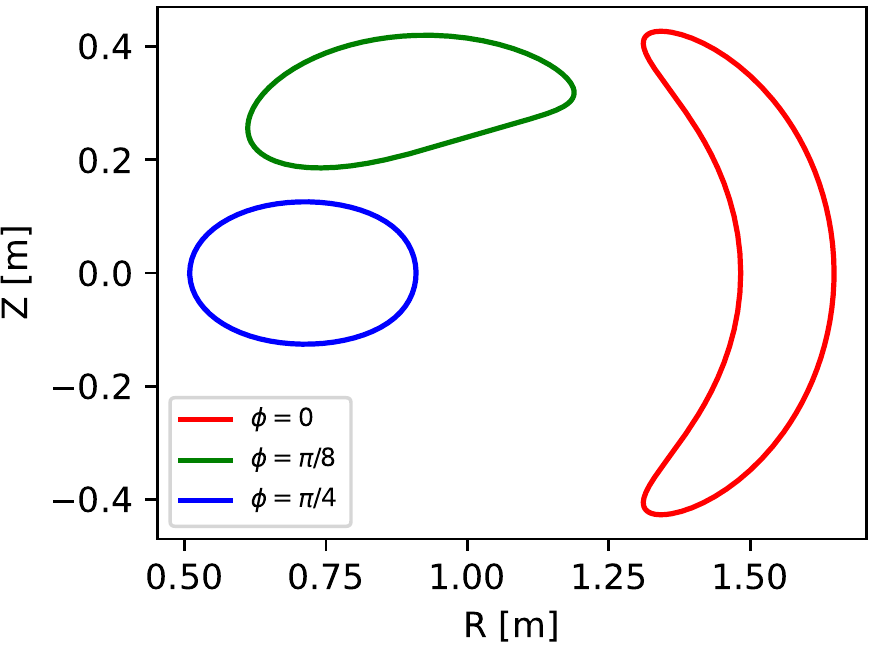}
\hspace{0.1in}
\includegraphics[width=3.1in]{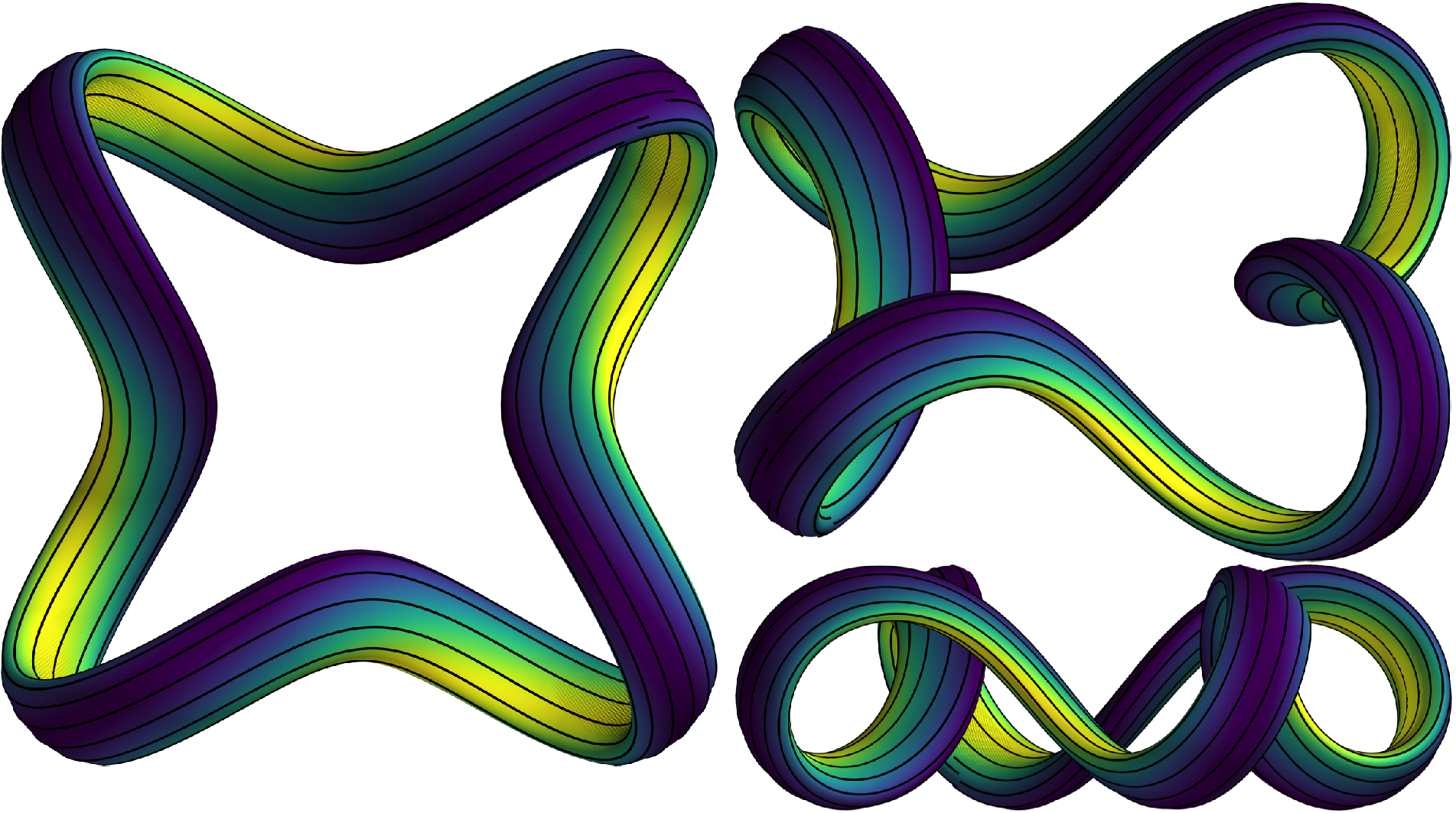}
  \caption{\label{fig:QH_nfp4_longAxis} 
A four-field-period quasi-helically symmetric stellarator generated from the near-axis method, with large ratio of magnetic axis length to major radius (12.0). Left: cross-sections. Right: The same configuration is shown from three perspectives. Color indicates the field strength, and black curves are field lines.
}
\end{figure*}

\begin{figure*}
\includegraphics[width=\columnwidth]{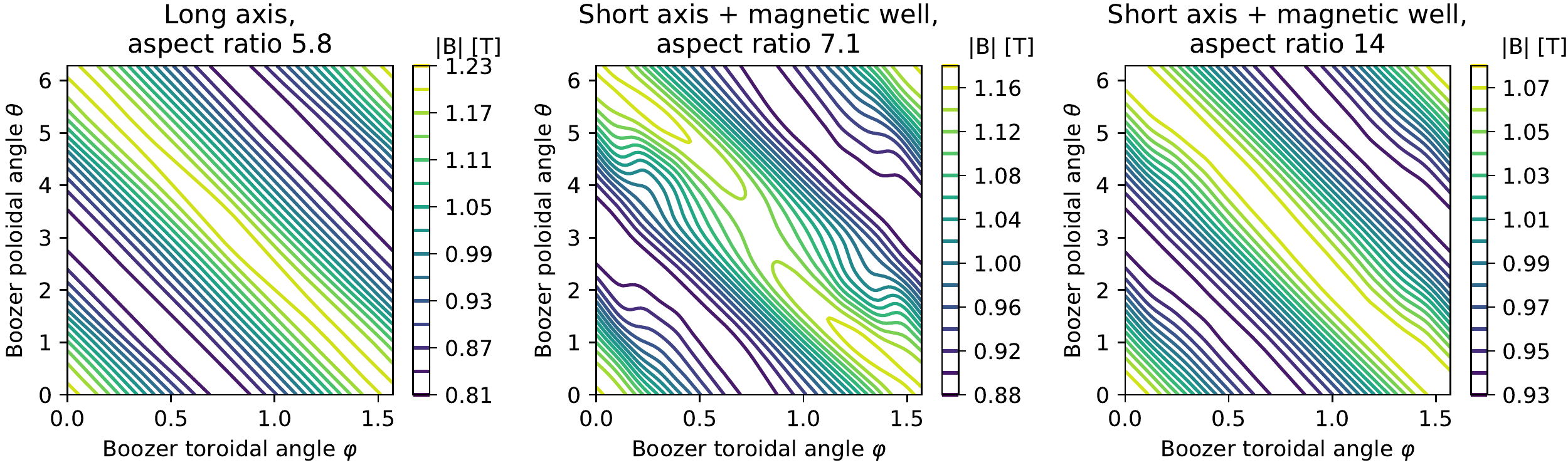}
  \caption{\label{fig:nfp4_QH_Boozer} 
  Magnetic field strength on the boundaries of the 
 four-field-period quasi-helically symmetric stellarators, computed by running a fully 3D fixed-boundary equilibrium calculation
inside the boundary constructed from the near-axis method.
The mostly straight diagonal contours confirm the QH symmetry. 
The configurations correspond to figures \ref{fig:QH_nfp4_longAxis} and \ref{fig:QH_nfp4_magwell}.
}
\end{figure*}

The first is a configuration with relatively long magnetic axis, $L / R_0 = 12$. This configuration is relatively far to the right on the band of four-field-period QH data in figure \ref{fig:landscape}. It is a vacuum field, and the only terms included in the optimization were $f_L$, $f_{\nabla}$, $f_{\nabla\nabla}$, and $f_{B2}$. 
The rotational transform is $\iota = 1.78$.
The plasma shape is shown in figure \ref{fig:QH_nfp4_longAxis} for $a=0.13$, corresponding to $A=5.8$. 
The axis shape of this configuration resembles the one in Fig.~2 of \cite{RodriguezLandscape}.
Given the boundary computed by the near-axis equations, the field strength inside is computed with DESC and displayed in figure \ref{fig:nfp4_QH_Boozer}, showing excellent QH symmetry.
\changed{This configuration has a magnetic hill.}

\begin{figure*}
  \centering
\includegraphics[width=1.6in]{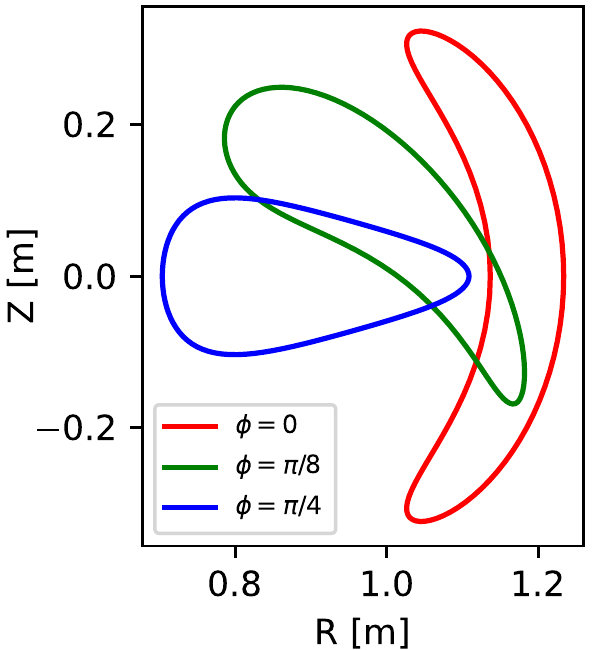}
\hspace{0.1in}
\includegraphics[width=3.4in]{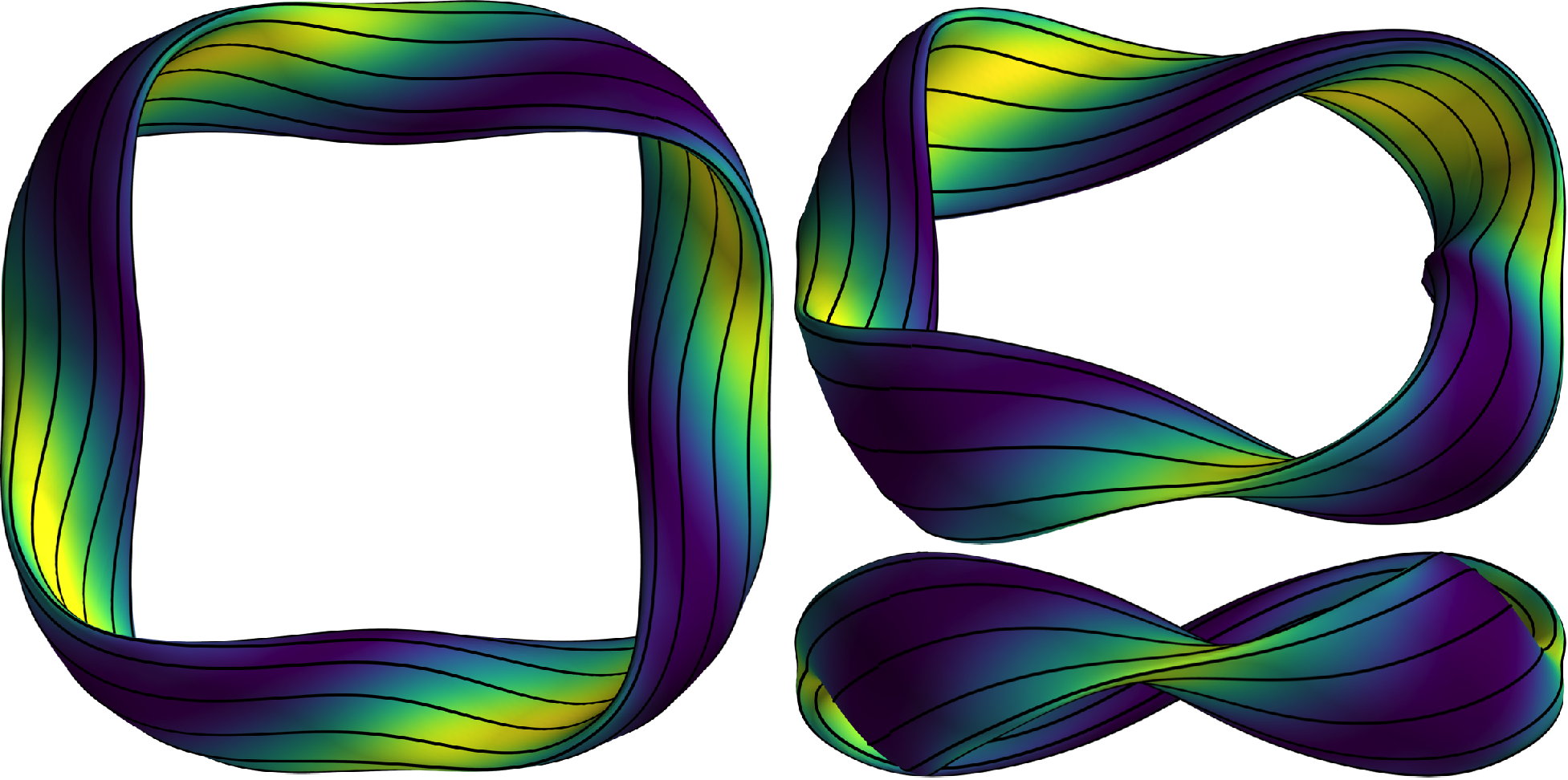}
  \caption{\label{fig:QH_nfp4_magwell} 
A four-field-period quasi-helically symmetric stellarator generated from the near-axis method, with magnetic well. Left: cross-sections. Right: The same configuration is shown from three perspectives. Color indicates the field strength, and black curves are field lines.
}
\end{figure*}

Next, we present a vacuum configuration with magnetic well, obtained by including the $f_{\mathrm{well}}$ term in the objective. This configuration has a magnetic axis length $L/R_0=7.0$, shorter than the previous configuration. 
The rotational transform is $\iota = 1.18$.
The plasma shape is shown in figure \ref{fig:QH_nfp4_magwell} for $a=0.13 R_0$, corresponding this time to $A=7.1$. 
This plasma shape is fairly similar to previous four-field-period QHs
\citep{HSX, KuBoozerQHS, Bader2020, LandremanPaul2022}.
The field strength inside this boundary computed by DESC is shown in figure \ref{fig:nfp4_QH_Boozer}. A QH pattern is apparent, though the deviations from QH symmetry are larger than in the previous configuration, even though $a$ is identical and $A$ is larger. This finding, that there is a significant trade-off between quasisymmetry and magnetic well, was also observed in \cite{LandremanPaul2022}. The quality of quasisymmetry can be improved to any desired degree by increasing the aspect ratio. This is shown in the right panel of figure \ref{fig:nfp4_QH_Boozer}, displaying the field strength from DESC when the plasma boundary is constructed for a value of $r$ that is half as large, giving $A=14$. For this higher aspect ratio the $B$ contours are significantly straighter.

\begin{figure*}
  \centering
\includegraphics[width=1.9in]{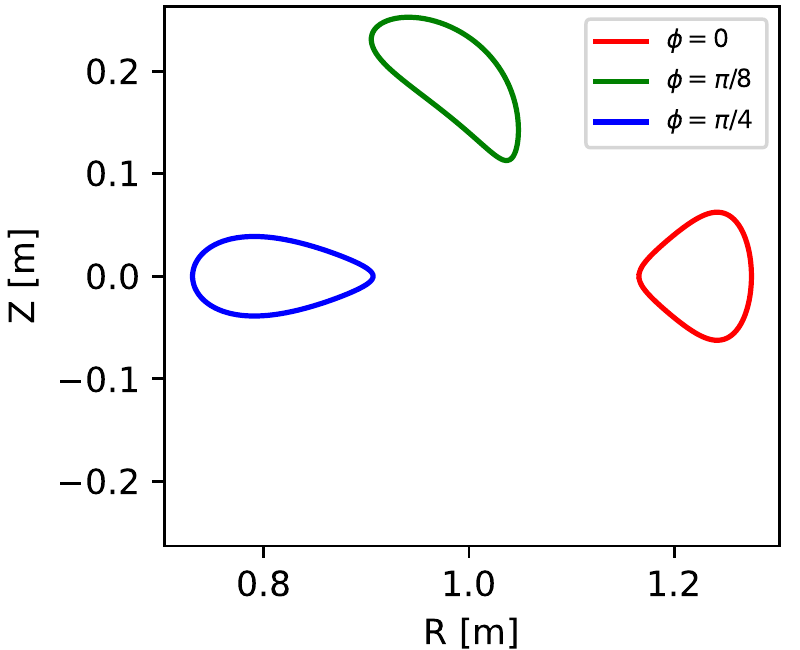}
\hspace{0.1in}
\includegraphics[width=3.2in]{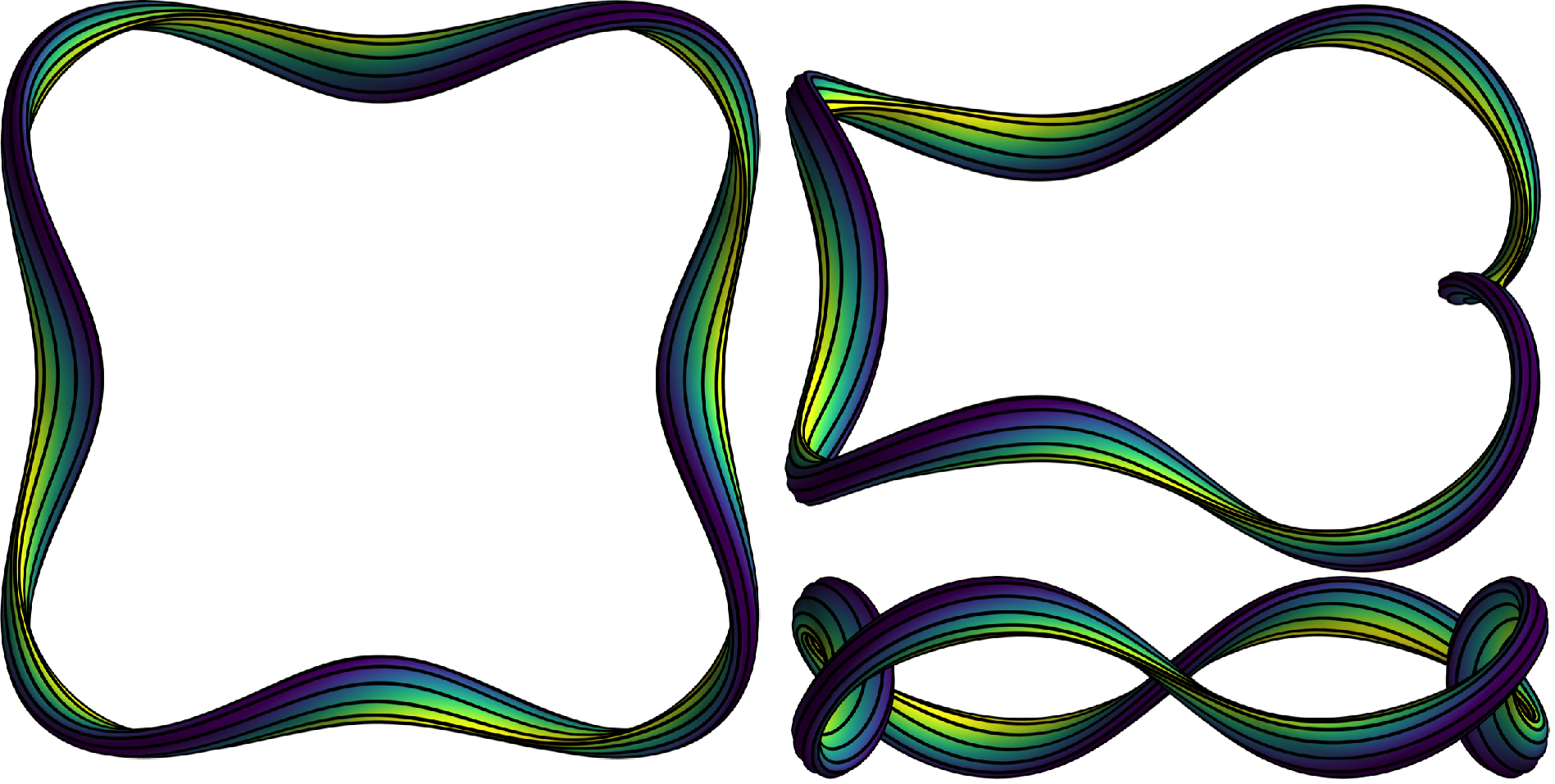}
  \caption{\label{fig:QH_nfp4_Mercier} 
A four-field-period quasi-helically symmetric stellarator generated from the near-axis method, with Mercier stability. This configuration is unique in having a reversed-triangularity cross-section in place of the usual bean-shaped cross-section. Left: cross-sections. Right: The same configuration is shown from three perspectives. Color indicates the field strength, and black curves are field lines.
}
\end{figure*}

\begin{figure*}
  \centering
\includegraphics[height=2.0in]{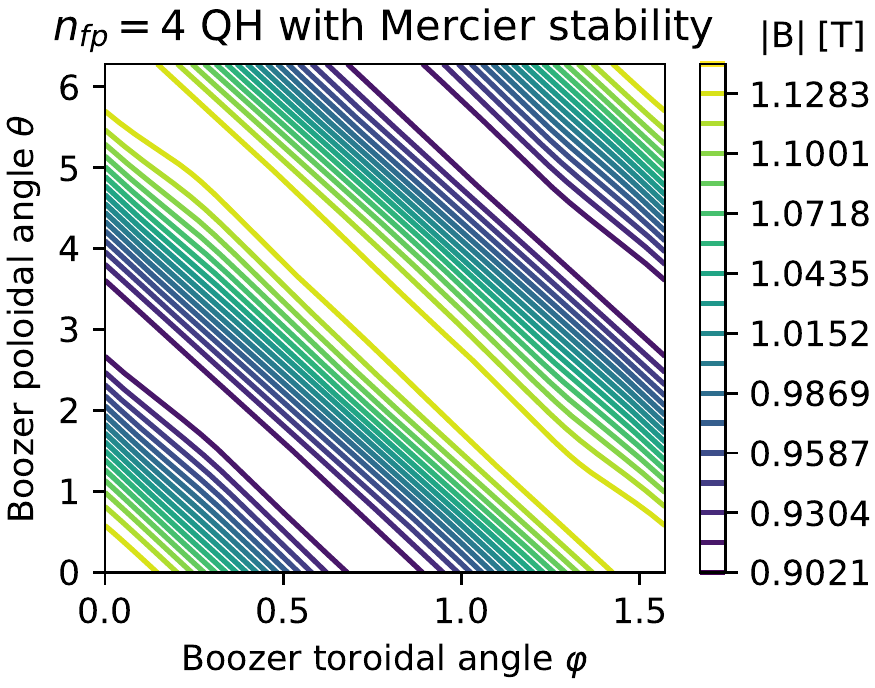}
  \caption{\label{fig:nfp4_QH_Mercier_Boozer} 
Magnetic field strength on the boundary of the 
 four-field-period quasi-helically symmetric stellarator with Mercier stability, computed by running a fully 3D fixed-boundary equilibrium calculation
inside the boundary constructed from the near-axis method.
The straight diagonal contours confirm the QH symmetry.
}
\end{figure*}

Finally, a four-field-period configuration with finite $\beta$ and Mercier stability is presented in figure \ref{fig:QH_nfp4_Mercier}.
For this configuration, 
a finite pressure gradient is included by setting $p_2=-10^6$ Pa$/$m$^2$, and
the $f_{\mathrm{Merc}}$ term is included in the objective. 
The rotational transform of the configuration is $\iota=1.60$.
Based on experience so far, including $f_{\mathrm{Merc}}$ in the optimization causes a substantial deterioration in the minimum aspect ratio, $R_0 / r_c$.
For this reason, a small minor radius is chosen for the plots, $a = 0.06 R_0$.
Almost all optimized stellarators have a ``bean-shaped'' cross-section, but 
figure \ref{fig:QH_nfp4_Mercier} shows that 
this configuration does not.
Instead, at the toroidal angle for which the major radius of the magnetic axis is maximized, this configuration has reversed triangularity.
This configuration also is unique in that it exhibits much stronger magnetic shear (computed from the fully 3D solution) than the other configurations in this paper.
We have observed similar solutions with Mercier stability also for $\nfp=3$.
The field strength on the finite aspect ratio boundary computed by DESC is shown in figure \ref{fig:nfp4_QH_Mercier_Boozer}, displaying the expected quasisymmetry.
In the future it would be valuable to further explore this unusual class of QH configurations that lack a bean-shaped cross-section.
Other important questions for future work are whether Mercier stability can be obtained with larger values of the minor radius $r_c$, and whether Mercier stability is in fact necessary or not in experiments.

\subsection{Quasi-helical symmetry with seven field periods}

\begin{figure*}
  \centering
\includegraphics[width=1.7in]{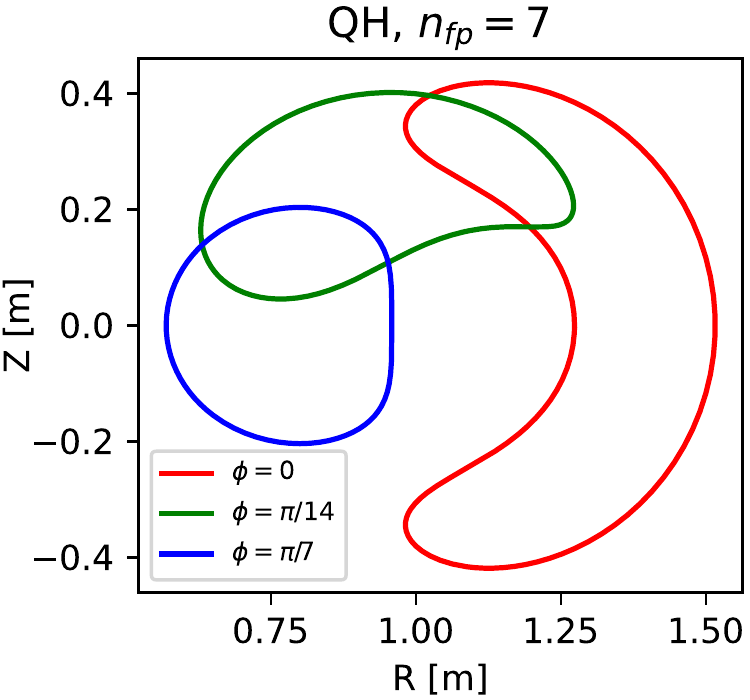}
\hspace{0.1in}
\includegraphics[width=3.4in]{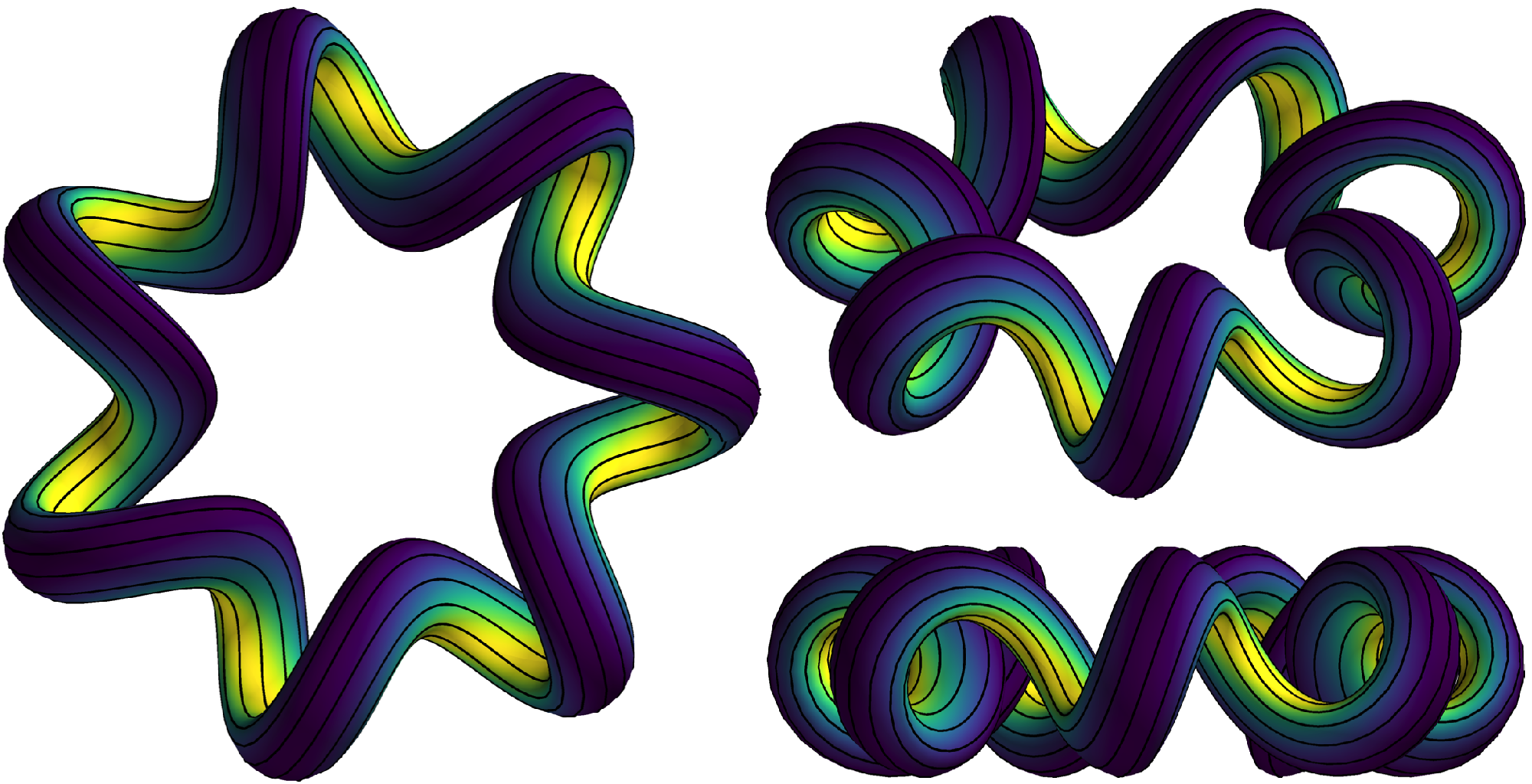}
  \caption{\label{fig:QH_nfp7} 
A seven-field-period quasi-helically symmetric stellarator generated from the near-axis method. Left: cross-sections. Right: The same configuration is shown from three perspectives. Color indicates the field strength, and black curves are field lines.
}
\end{figure*}

To our knowledge, in previously reported QH configurations, the highest number of field periods has been 6 \citep{NuhrenbergZille}. With the near-axis method, as already mentioned, QH solutions were found also for larger numbers of field periods, passing the filters also for $\nfp=7$ and 8. A 7-field-period configuration is shown in figure \ref{fig:QH_nfp7},
for $a / R_0 = 0.15$. 
This configuration is a vacuum field with very large rotational transform, $\iota \approx 3.65$.
A target axis length $L_* / R_0 = 14$ was used.
As with the $\nfp=2$ QH configuration, it takes a large number of Fourier modes to represent this configuration in cylindrical coordinates.
This can be understood from the unusual shaping, with sections of the plasma column that are nearly vertical.

\begin{figure*}
  \centering
\includegraphics[height=2.0in]{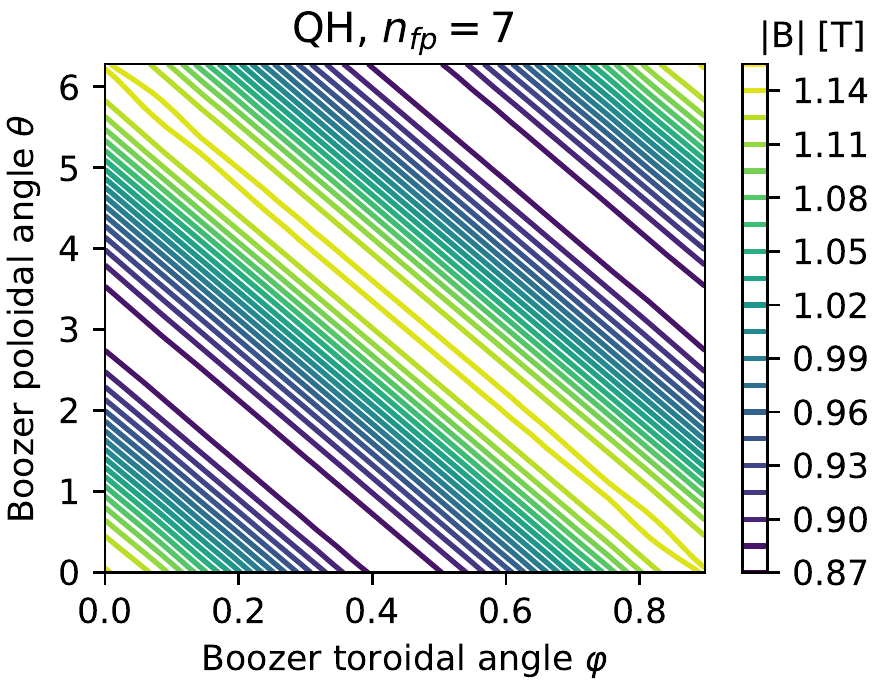}
  \caption{\label{fig:nfp7_QH_Boozer} 
Magnetic field strength on the $a/R_0=0.05$ surface of the 
 seven-field-period quasi-helically symmetric stellarator, computed by running a fully 3D fixed-boundary equilibrium calculation
inside the boundary constructed from the near-axis method.
The straight diagonal contours confirm the QH symmetry.
}
\end{figure*}

Figure \ref{fig:nfp7_QH_Boozer} shows the field strength computed with DESC for the surface with $a/R_0=0.05$. 
The straight $B$ contours in the figure confirm the good QH symmetry.

\section{Discussion and conclusions}
\label{sec:conclusions}

In this work we have demonstrated a method to rapidly compute approximately quasisymmetric stellarator equilibria, and to map out the space of quasisymmetric configurations. 
The approach is based on expanding the relevant equations about the magnetic axis, and applying optimization to the reduced equations.
Optimization is not required to obtain quasisymmetry when using this expansion, since it can
be imposed directly \changed{(to $O(r)$)} in a neighborhood of the axis. However it is useful to apply optimization in practice to the axis shape and other near-axis parameters, to increase the range of minor radius over which the expansion is accurate.
Optimization can also be used to achieve other potentially desirable properties such as magnetic well or a desired rotational transform.
A large number of diagnostic quantities can be computed directly within the near-axis expansion and included in the objective function.
Due to the reduction of the equations by the expansion, a complete optimization takes only on the order of a cpu-second. Therefore it is feasible to carry out wide scans over parameter space.

From the parameter scans shown in figure \ref{fig:landscape}, 
several previous observations about quasisymmetric stellarators are reproduced,
and some new observations can be made.
QA solutions are best obtained at $\nfp=2$, with some marginal solutions also for $\nfp=3$, and are limited to $\iota<1$.
QH solutions have  $\iota > 0.9$.
As recently observed by \cite{RodriguezLandscape}, QH and QA solutions exist in continuous bands along which the axis length varies.
Along the QA bands, $\iota$ varies significantly, whereas $\iota$ varies more weakly along the QH bands.
QH solutions also exist with many possible values of $\nfp$, including as few as two. 

\changed{In the remainder of this section, we list some of the many directions for future work. First,}
more could be done to explore patterns in the database of configurations from section \ref{sec:scan}. Using other configuration properties besides $\iota$ and axis length, the data may be separable into clusters differently, such as configurations with vs without a bean cross-section (e.g. figure \ref{fig:QH_nfp4_Mercier}). It would also be valuable to try to understand patterns in the data, such as the fact that QA solutions seem limited to $\nfp=2$ and 3, by applying analytic methods to the underlying Garren-Boozer equations.
Structures in the space of configurations were recently explored using a different method in \cite{RodriguezLandscape}, and hopefully connections could be drawn between that work and the methods here.

Some of the QH solutions here, such as those in figures \ref{fig:nfp2_QH} and \ref{fig:QH_nfp7}, may not have been seen previously since they require many Fourier modes to represent using the usual boundary shape representation in cylindrical coordinates. It may therefore be valuable to develop near-axis and 3D MHD equilibrium codes that can use other coordinate systems, such as is being pursued with the code GVEC \citep{GVEC}. 
It would also be advantageous to modify the workflow used here so the surfaces are constructed using a poloidal angle other than the Boozer $\theta$, an angle in which the Fourier spectrum of the surface is more compressed.

We also find in the scans that there is a significant trade-off between the accuracy of quasisymmetry versus magnetic well (relevant for low $\beta$) or Mercier stability (relevant at finite $\beta$.) This finding motivates further work on nonlinear MHD stability, to assess whether these measures of linear stability are in fact necessary constraints to impose on a design, or whether they can be relaxed.

Compared to the unconstrained local optimizations used in this work,
other optimization methods could be applied to the near-axis model in the future.
Algorithms for  optimization with constraints could be used instead of the unconstrained approach with penalty terms used here. 
Also, global algorithms could be applied, since there is no guarantee that the scans here have found all global optima.

There are many other directions for future work.
One important question is how to include the bootstrap current in the near-axis model,
given the limited freedom in the current profile shape at order $O(r^2)$.
Second, the methods here could be further developed for quasi-isodynamic configurations, building on the work  in \cite{PaperIII,JorgeQI,Mata}.
Finally,
there is potential for using the geometry relevant to the gyrokinetic equation and ballooning stability, computed from the near-axis quantities by \cite{JorgeGKGeometry}. Properties of gyrokinetic or ballooning modes 
could potentially be targeted in the optimizations.


\section*{Acknowledgements}

Conversations about the near-axis expansion with Rogerio Jorge and Eduardo Rodriguez are gratefully acknowledged. Assistance with the DESC code was provided by Daniel Dudt, Rory Conlin, and Dario Panici.

\section*{Funding}

This work was supported by the U.S. Department of Energy, Office of Science, Office of Fusion Energy Science, under award number DE-FG02-93ER54197.

\section*{Declaration of interests}

The author reports no conflict of interest.

\section*{Data availability statement}

The data that support the findings of this study are openly available in Zenodo at \url{https://doi.org/10.5281/zenodo.7108893}

\section*{Author ORCID}

M. Landreman, \url{https://orcid.org/0000-0002-7233-577X}


\bibliographystyle{jpp}

\bibliography{Near_axis_quasisymmetry_landscape}

\begin{thebibliography}{45}
\expandafter\ifx\csname natexlab\endcsname\relax\def\natexlab#1{#1}\fi
\def\au#1{#1} \def\ed#1{#1} \def\yr#1{#1}\def\at#1{#1}\def\jt#1{\textit{#1}}
  \def\bt#1{#1}\def\bvol#1{\textbf{#1}} \def\vol#1{#1} \def\pg#1{#1}
  \def\publ#1{#1}\def\arxiv#1{#1}\def\org#1{#1}\def\st#1{\textit{#1}}

\bibitem[Anderson {\em et~al.\/}(1995)Anderson, Almagri, Anderson, Matthews,
  Talmadge \& Shohet]{HSX}
{\sc \au{Anderson, F Simon~B}, \au{Almagri, Abdulgader~F}, \au{Anderson,
  David~T}, \au{Matthews, Peter~G}, \au{Talmadge, Joseph~N} \& \au{Shohet,
  J~Leon}} \yr{1995}  \at{The helically symmetric experiment, {(HSX)} goals,
  design and status}.  \jt{Fusion Technology}  \bvol{27},  \pg{273}.

\bibitem[Bader {\em et~al.\/}(2021)Bader, Anderson, Drevlak, Faber, Hegna,
  Henneberg, Landreman, Schmitt, Suzuki \& Ware]{Bader2021}
{\sc \au{Bader, A}, \au{Anderson, DT}, \au{Drevlak, M}, \au{Faber, BJ},
  \au{Hegna, CC}, \au{Henneberg, S}, \au{Landreman, M}, \au{Schmitt, JC},
  \au{Suzuki, Y} \& \au{Ware, A}} \yr{2021}  \at{Modeling of energetic particle
  transport in optimized stellarators}.  \jt{Nuclear Fusion}  \bvol{61},
  \pg{116060}.

\bibitem[Bader {\em et~al.\/}(2019)Bader, Drevlak, Anderson, Faber, Hegna,
  Likin, Schmitt \& Talmadge]{Bader2019}
{\sc \au{Bader, Aaron}, \au{Drevlak, M}, \au{Anderson, DT}, \au{Faber, BJ},
  \au{Hegna, CC}, \au{Likin, KM}, \au{Schmitt, JC} \& \au{Talmadge, JN}}
  \yr{2019}  \at{Stellarator equilibria with reactor relevant energetic
  particle losses}.  \jt{J. Plasma Phys.}  \bvol{85}.

\bibitem[Bader {\em et~al.\/}(2020)Bader, Faber, Schmitt, Anderson, Drevlak,
  Duff, Frerichs, Hegna, Kruger, Landreman {\em et~al.\/}]{Bader2020}
{\sc \au{Bader, A}, \au{Faber, BJ}, \au{Schmitt, JC}, \au{Anderson, DT},
  \au{Drevlak, M}, \au{Duff, JM}, \au{Frerichs, H}, \au{Hegna, CC}, \au{Kruger,
  TG}, \au{Landreman, M} \& \au{others}} \yr{2020}  \at{Advancing the physics
  basis for quasi-helically symmetric stellarators}.  \jt{J. Plasma Phys.}
  \bvol{86}.

\bibitem[Boozer(1983)]{Boozer1983}
{\sc \au{Boozer, Allen~H}} \yr{1983}  \at{Transport and isomorphic equilibria}.
   \jt{Phys. Fluids}  \bvol{26},  \pg{496}.

\bibitem[Boozer(2020)]{Boozer2020}
{\sc \au{Boozer, Allen~H}} \yr{2020}  \at{Why carbon dioxide makes stellarators
  so important}.  \jt{Nucl. Fusion}  \bvol{60},  \pg{065001}.

\bibitem[{Camacho Mata} {\em et~al.\/}(2022){Camacho Mata}, Plunk \&
  Jorge]{Mata}
{\sc \au{{Camacho Mata}, K}, \au{Plunk, G~G} \& \au{Jorge, R}} \yr{2022}
  \at{Direct construction of stellarator-symmetric quasi-isodynamic magnetic
  configurations}.  \jt{J. Plasma Phys.}  \bvol{88},  \pg{905880503}.

\bibitem[Conlin {\em et~al.\/}(2022)Conlin, Dudt, Panici \& Kolemen]{Desc2}
{\sc \au{Conlin, Rory}, \au{Dudt, Daniel~W}, \au{Panici, Dario} \& \au{Kolemen,
  Egemen}} \yr{2022}  \at{The desc stellarator code suite part {II}:
  Perturbation and continuation methods}.  \jt{arXiv preprint arXiv:2203.15927}
  .

\bibitem[{de Aguilera} {\em et~al.\/}(2015){de Aguilera}, Castejon, Ascasibar,
  Blanco, {de la Cal}, Hidalgo, Liu, Lopez-Fraguas, Medina, Ochando, Pastor,
  Pedrosa, {van Milligen}, Velasco \& {the TJ-II team}]{Aguilera}
{\sc \au{{de Aguilera}, A~M}, \au{Castejon, F}, \au{Ascasibar, E}, \au{Blanco,
  E}, \au{{de la Cal}, E}, \au{Hidalgo, C}, \au{Liu, B}, \au{Lopez-Fraguas, A},
  \au{Medina, F}, \au{Ochando, M~A}, \au{Pastor, I}, \au{Pedrosa, M~A},
  \au{{van Milligen}, B}, \au{Velasco, J~L} \& \au{{the TJ-II team}}} \yr{2015}
   \at{Magnetic well scan and confinement in the {TJ-II} stellarator}.
  \jt{Nucl. Fusion}  \bvol{55},  \pg{113014}.

\bibitem[D'haeseleer {\em et~al.\/}(2012)D'haeseleer, Hitchon, Callen \&
  Shohet]{Dhaeseleer}
{\sc \au{D'haeseleer, William~D}, \au{Hitchon, William~NG}, \au{Callen,
  James~D} \& \au{Shohet, J~Leon}} \yr{2012} {\em Flux coordinates and magnetic
  field structure: a guide to a fundamental tool of plasma theory\/}.
  \publ{Springer Science \& Business Media}.

\bibitem[Dudt {\em et~al.\/}(2022)Dudt, Conlin, Panici \& Kolemen]{Desc3}
{\sc \au{Dudt, Daniel}, \au{Conlin, Rory}, \au{Panici, Dario} \& \au{Kolemen,
  Egemen}} \yr{2022}  \at{The desc stellarator code suite part {III}:
  Quasi-symmetry optimization}.  \jt{arXiv preprint arXiv:2204.00078} .

\bibitem[Dudt \& Kolemen(2020)]{Desc0}
{\sc \au{Dudt, DW} \& \au{Kolemen, E}} \yr{2020}  \at{Desc: A stellarator
  equilibrium solver}.  \jt{Phys. Plasmas}  \bvol{27},  \pg{102513}.

\bibitem[Galassi(2009)]{GSL}
{\sc \au{Galassi, {M, \emph{et al}}}} \yr{2009} {\em GNU Scientific Library
  Reference Manual (3rd Ed.)\/}.

\bibitem[Garren \& Boozer(1991{\natexlab{{\em a\/}}})]{GB2}
{\sc \au{Garren, D~A} \& \au{Boozer, A~H}} \yr{1991{\natexlab{{\em a\/}}}}
  \at{Existence of quasihelically symmetric stellarators}.  \jt{Phys. Fluids B}
   \bvol{3},  \pg{2822}.

\bibitem[Garren \& Boozer(1991{\natexlab{{\em b\/}}})]{GB1}
{\sc \au{Garren, D~A} \& \au{Boozer, A~H}} \yr{1991{\natexlab{{\em b\/}}}}
  \at{Magnetic field strength of toroidal plasma equilibria}.  \jt{Phys. Fluids
  B}  \bvol{3},  \pg{2805}.

\bibitem[Geiger {\em et~al.\/}(2004)Geiger, Weller, Zarnstorff, N\"{u}hrenberg,
  Werner, Kolesnichenko \& {the W7-AS team}]{Geiger}
{\sc \au{Geiger, J~E}, \au{Weller, A}, \au{Zarnstorff, M~C},
  \au{N\"{u}hrenberg, C}, \au{Werner, A}, \au{Kolesnichenko, Y~I} \& \au{{the
  W7-AS team}}} \yr{2004}  \at{Equilibrium and stability of high-$\beta$
  plasmas in {W}endelstein 7-{AS}}.  \jt{Fusion Sci. Tech.}  \bvol{46},
  \pg{13}.

\bibitem[Giuliani {\em et~al.\/}(2022)Giuliani, Wechsung, Landreman, Stadler \&
  Cerfon]{GiulianiSurface}
{\sc \au{Giuliani, Andrew}, \au{Wechsung, Florian}, \au{Landreman, Matt},
  \au{Stadler, Georg} \& \au{Cerfon, Antoine}} \yr{2022}  \at{Direct
  computation of magnetic surfaces in {B}oozer coordinates and coil
  optimization for quasi-symmetry}.  \jt{J. Plasma Phys.}  \bvol{88},
  \pg{905880401}.

\bibitem[Helander(2014)]{HelanderReview}
{\sc \au{Helander, Per}} \yr{2014}  \at{Theory of plasma confinement in
  non-axisymmetric magnetic fields}.  \jt{Reports on Progress in Physics}
  \bvol{77},  \pg{087001}.

\bibitem[Hirshman \& Whitson(1983)]{VMEC1983}
{\sc \au{Hirshman, S~P} \& \au{Whitson, J~C}} \yr{1983}  \at{Steepest-descent
  moment method for three-dimensional magnetohydrodynamic equilibria}.
  \jt{Phys. Fluids}  \bvol{26},  \pg{3553}.

\bibitem[Jorge \& Landreman(2021)]{JorgeGKGeometry}
{\sc \au{Jorge, R} \& \au{Landreman, M}} \yr{2021}  \at{The use of near-axis
  magnetic fields for stellarator turbulence simulations}.  \jt{Plasma Phys.
  Controlled Fusion}  \bvol{63},  \pg{014001}.

\bibitem[Jorge {\em et~al.\/}(2022)Jorge, Plunk, Drevlak, Landreman, Lobsien,
  Mata \& Helander]{JorgeQI}
{\sc \au{Jorge, R}, \au{Plunk, GG}, \au{Drevlak, M}, \au{Landreman, M},
  \au{Lobsien, J-F}, \au{Mata, K~Camacho} \& \au{Helander, P}} \yr{2022}  \at{A
  single-field-period quasi-isodynamic stellarator}.  \jt{J. Plasma Phys.}
  \bvol{88},  \pg{175880504}.

\bibitem[Jorge {\em et~al.\/}(2020)Jorge, Sengupta \& Landreman]{Jorge2021}
{\sc \au{Jorge, R}, \au{Sengupta, W} \& \au{Landreman, M}} \yr{2020}
  \at{{Near-Axis Expansion of Stellarator Equilibrium at Arbitrary Order in the
  Distance to the Axis}}.  \jt{J. Plasma Phys.}  \bvol{86},  \pg{905860106}.

\bibitem[Ku \& Boozer(2011)]{KuBoozerQHS}
{\sc \au{Ku, LP} \& \au{Boozer, AH}} \yr{2011}  \at{New classes of
  quasi-helically symmetric stellarators}.  \jt{Nucl. Fusion}  \bvol{51},
  \pg{013004}.

\bibitem[Landreman(2021)]{FiguresOfMerit}
{\sc \au{Landreman, M}} \yr{2021}  \at{Figures of merit for stellarators near
  the magnetic axis}.  \jt{J. Plasma Phys.}  \bvol{87},  \pg{905870112}.

\bibitem[Landreman(2022)]{zenodoData}
{\sc \au{Landreman, M}} \yr{2022}  \at{Dataset on {Zenodo},
  \url{https://doi.org/10.5281/zenodo.7108893}} .

\bibitem[Landreman \& Jorge(2020)]{LandremanMercier}
{\sc \au{Landreman, M} \& \au{Jorge, R}} \yr{2020}  \at{Mercier stability of
  stellarators near the magnetic axis}.  \jt{J. Plasma Phys.}  \bvol{86},
  \pg{905860510}.

\bibitem[Landreman \& Paul(2022)]{LandremanPaul2022}
{\sc \au{Landreman, M} \& \au{Paul, E}} \yr{2022}  \at{Magnetic fields with
  precise quasisymmetry for plasma confinement}.  \jt{Phys. Rev. Lett.}
  \bvol{128},  \pg{035001}.

\bibitem[Landreman \& Sengupta(2018)]{PaperI}
{\sc \au{Landreman, M} \& \au{Sengupta, W}} \yr{2018}  \at{{Direct construction
  of optimized stellarator shapes. I. Theory in cylindrical coordinates}}.
  \jt{J. Plasma Phys.}  \bvol{84},  \pg{905840616}.

\bibitem[Landreman \& Sengupta(2019)]{r2GarrenBoozer}
{\sc \au{Landreman, M} \& \au{Sengupta, W}} \yr{2019}  \at{{Constructing
  stellarators with quasisymmetry to high order}}.  \jt{J. Plasma Phys.}
  \bvol{85},  \pg{905850608}.

\bibitem[Landreman {\em et~al.\/}(2019)Landreman, Sengupta \& Plunk]{PaperII}
{\sc \au{Landreman, M}, \au{Sengupta, W} \& \au{Plunk, G~G}} \yr{2019}
  \at{{Direct construction of optimized stellarator shapes. II. Numerical
  quasisymmetric solutions}}.  \jt{J. Plasma Phys.}  \bvol{85},
  \pg{905850103}.

\bibitem[Liu {\em et~al.\/}(2018)Liu, Shimizu, Isobe, Okamura, Nishimura,
  Suzuki, Xu, Zhang, Liu, Huang {\em et~al.\/}]{CFQS}
{\sc \au{Liu, Haifeng}, \au{Shimizu, Akihiro}, \au{Isobe, Mitsutaka},
  \au{Okamura, Shoichi}, \au{Nishimura, Shin}, \au{Suzuki, Chihiro}, \au{Xu,
  Yuhong}, \au{Zhang, Xin}, \au{Liu, Bing}, \au{Huang, Jie} \& \au{others}}
  \yr{2018}  \at{Magnetic configuration and modular coil design for the
  {Chinese First Quasi-Axisymmetric Stellarator}}.  \jt{Plasma and Fusion
  Research}  \bvol{13},  \pg{3405067}.

\bibitem[Maurer {\em et~al.\/}(2020)Maurer, Navarro, Dannert, Restelli,
  Hindenlang, G{\"o}rler, Told, Jarema, Merlo \& Jenko]{GVEC}
{\sc \au{Maurer, Maurice}, \au{Navarro, A~Ba{\~n}{\'o}n}, \au{Dannert, Tilman},
  \au{Restelli, Marco}, \au{Hindenlang, Florian}, \au{G{\"o}rler, Tobias},
  \au{Told, Daniel}, \au{Jarema, Denis}, \au{Merlo, Gabriele} \& \au{Jenko,
  Frank}} \yr{2020}  \at{Gene-3d: A global gyrokinetic turbulence code for
  stellarators}.  \jt{J. Comp. Phys.}  \bvol{420},  \pg{109694}.

\bibitem[Mercier(1964)]{Mercier1964}
{\sc \au{Mercier, C}} \yr{1964}  \at{Equilibrium and stability of a toroidal
  magnetohydrodynamic system in the neighbourhood of a magnetic axis}.
  \jt{Nucl. Fusion}  \bvol{4},  \pg{213}.

\bibitem[N\"{u}hrenberg \& Zille(1988)]{NuhrenbergZille}
{\sc \au{N\"{u}hrenberg, J} \& \au{Zille, R}} \yr{1988}  \at{Quasi-helically
  symmetric toroidal stellarators}.  \jt{Phys. Lett. A}  \bvol{129},  \pg{113}.

\bibitem[Panici {\em et~al.\/}(2022)Panici, Conlin, Dudt \& Kolemen]{Desc1}
{\sc \au{Panici, Dario}, \au{Conlin, Rory}, \au{Dudt, Daniel~W} \& \au{Kolemen,
  Egemen}} \yr{2022}  \at{The desc stellarator code suite part {I}: Quick and
  accurate equilibria computations}.  \jt{arXiv preprint arXiv:2203.17173} .

\bibitem[Paul {\em et~al.\/}(2022)Paul, Bhattacharjee, Landreman, Alex, Velasco
  \& Nies]{Paul2022}
{\sc \au{Paul, EJ}, \au{Bhattacharjee, A}, \au{Landreman, M}, \au{Alex, D},
  \au{Velasco, JL} \& \au{Nies, R}} \yr{2022}  \at{Energetic particle loss
  mechanisms in reactor-scale equilibria close to quasisymmetry}.  \jt{Nucl.
  Fusion}  \bvol{62},  \pg{126054}.

\bibitem[Plunk {\em et~al.\/}(2019)Plunk, Landreman \& Helander]{PaperIII}
{\sc \au{Plunk, G~G}, \au{Landreman, M} \& \au{Helander, P}} \yr{2019}
  \at{{Direct construction of optimized stellarator shapes. III. Omnigenity
  near the magnetic axis}}.  \jt{J. Plasma Phys.}  \bvol{85},  \pg{905850602}.

\bibitem[Rodriguez(2022)]{RodriguezThesis}
{\sc \au{Rodriguez, E}} \yr{2022}  \at{Quasisymmetry}. PhD thesis, Princeton
  University.

\bibitem[Rodriguez {\em et~al.\/}(2022{\natexlab{{\em a\/}}})Rodriguez,
  Sengupta \& Bhattacharjee]{RodriguezPhases}
{\sc \au{Rodriguez, E}, \au{Sengupta, W} \& \au{Bhattacharjee, A}}
  \yr{2022{\natexlab{{\em a\/}}}}  \at{Phases and phase-transitions in
  quasisymmetric configuration space}.  \jt{Plasma Phys. Controlled Fusion}
  \bvol{64},  \pg{105006}.

\bibitem[Rodriguez {\em et~al.\/}(2022{\natexlab{{\em b\/}}})Rodriguez,
  Sengupta \& Bhattacharjee]{RodriguezLandscape}
{\sc \au{Rodriguez, Eduardo}, \au{Sengupta, Wrick} \& \au{Bhattacharjee,
  Amitava}} \yr{2022{\natexlab{{\em b\/}}}}  \at{Topology-mediated approach to
  the design of quasisymmetric stellarators}.  \jt{arXiv preprint
  arXiv:2204.10234} .

\bibitem[Rodr{\'\i}guez {\em et~al.\/}(2022)Rodr{\'\i}guez, Sengupta \&
  Bhattacharjee]{Rodriguez2022weakly}
{\sc \au{Rodr{\'\i}guez, Eduardo}, \au{Sengupta, Wrick} \& \au{Bhattacharjee,
  Amitava}} \yr{2022}  \at{Weakly quasisymmetric near-axis solutions to all
  orders}.  \jt{Physics of Plasmas}  \bvol{29},  \pg{012507}.

\bibitem[Solov’ev \& Shafranov(1970)]{Solovev}
{\sc \au{Solov’ev, LS} \& \au{Shafranov, VD}} \yr{1970}  \at{Closed magnetic
  configurations for plasma confinement}.  \jt{Reviews of Plasma Physics}
  \bvol{5},  \pg{1--247}.

\bibitem[Spitzer(1958)]{Spitzer58}
{\sc \au{Spitzer, L}} \yr{1958}  \at{The stellarator concept}.  \jt{Phys.
  Fluids}  \bvol{1},  \pg{253}.

\bibitem[Watanabe {\em et~al.\/}(2005)Watanabe, Sakakibara, Narushima, Funaba,
  Narihara, Tanaka, Yamaguchi, Toi, Ohdachi, Kaneko, Yamada, Suzuki, Cooper,
  Murakami, Nakajima, Yamada, Kawahata, Tokuzawa, Komori \& {the LHD
  experimental group}]{Watanabe}
{\sc \au{Watanabe, K.~Y.}, \au{Sakakibara, S.}, \au{Narushima, Y.}, \au{Funaba,
  H.}, \au{Narihara, K.}, \au{Tanaka, K.}, \au{Yamaguchi, T.}, \au{Toi, K.},
  \au{Ohdachi, S.}, \au{Kaneko, O.}, \au{Yamada, H.}, \au{Suzuki, Y.},
  \au{Cooper, W.~A.}, \au{Murakami, S.}, \au{Nakajima, N.}, \au{Yamada, I.},
  \au{Kawahata, K.}, \au{Tokuzawa, T.}, \au{Komori, A.} \& \au{{the LHD
  experimental group}}} \yr{2005}  \at{Effects of global {MHD} instability on
  operational high beta-regime in {LHD}}.  \jt{Nucl. Fusion}  \bvol{45},
  \pg{1247}.

\bibitem[Weller {\em et~al.\/}(2006)Weller, Sakakibara, Watanabe, Toi, Geiger,
  Zarnstorff, Hudson, Reiman, Werner, N\"{u}hrenberg, Ohdachi, Suzuki, Yamada,
  {the W7-AS team} \& {the LHD team}]{Weller2006}
{\sc \au{Weller, A}, \au{Sakakibara, S}, \au{Watanabe, K~Y}, \au{Toi, K},
  \au{Geiger, J}, \au{Zarnstorff, M~C}, \au{Hudson, S~R}, \au{Reiman, A},
  \au{Werner, A}, \au{N\"{u}hrenberg, C}, \au{Ohdachi, S}, \au{Suzuki, Y},
  \au{Yamada, H}, \au{{the W7-AS team}} \& \au{{the LHD team}}} \yr{2006}
  \at{Significance of {MHD} effects in stellarator confinement}.  \jt{Fusion
  Sci. Tech.}  \bvol{50},  \pg{158}.

\end{thebibliography}

\end{document}